\documentclass[reprint,floatfix,amsmath,amssymb,aps,prd]{revtex4-2}

\usepackage{amsmath,amssymb,amsfonts}
\usepackage{bm}
\usepackage{graphicx}
\graphicspath{{./}{figures/}{Figures/}{./figures/}}
\usepackage{braket}
\usepackage{hyperref}
\usepackage[normalem]{ulem}
\usepackage[caption=false]{subfig}
\DeclareMathOperator{\sech}{sech}

\hypersetup{
	colorlinks=true,
	linkcolor=blue,
	citecolor=blue,
	urlcolor=blue,
}

\begin{document}
	
\title{Entanglement probes of gravitational Kaluza--Klein spectra: signal hierarchy and model discrimination}
	
\author{Yi Zhong}
\email{zhongy@hnu.edu.cn}
\affiliation{School of Physics and Electronics, Hunan University, Changsha 410082, China}
\affiliation{Hunan Provincial Key Laboratory of High-Energy Scale Physics and Applications, Changsha 410082, China}
	
\author{Tao-Tao Sui}
\email{suitaotao@aust.edu.cn}
\affiliation{Center for Fundamental Physics, School of Mechanics and Photoelectric Physics, Anhui University of Science and Technology, Huainan, Anhui 232001, China}
	
\author{Ke Yang}
\email{keyang@swu.edu.cn, corresponding author}
\affiliation{School of Physical Science and Technology, Southwest University, Chongqing 400715, China}

\begin{abstract}
Quantum-gravity-induced entanglement of masses (QGEM) provides a phase-sensitive probe of extra-dimensional corrections to the Newtonian potential at submillimeter separations. We compare three representative Kaluza--Klein spectral scenarios: the Randall--Sundrum II (RSII) and Arkani-Hamed--Dimopoulos--Dvali (ADD) models, and the case of a gapped continuum modeled by a P\"oschl--Teller potential. We evaluate the entangling phase, concurrence, and normalized phase-response profiles over $d=40$--$80\,\mu\mathrm{m}$ using representative benchmark parameters guided by current short-range gravity tests. In this range, the signal exhibits a stable hierarchy: ADD $>$ gapped $>$ RSII. For conservative experimental parameters, the ADD signal surpasses the nominal entanglement threshold at smaller separations, whereas the gapped benchmark is resolvable only at the lower end of the window, and RSII remains below resolution. In a more optimistic near-term scenario, all three spectral signatures comfortably exceed the threshold. We further show that normalized distance scans of the phase response clearly separate the RSII benchmark from the ADD and gapped cases, whereas ADD and the gapped continuum remain nearly indistinguishable in normalized  profile. QGEM phase observables therefore provide a complementary discriminator of Kaluza--Klein spectral structure at submillimeter scales.
\end{abstract}

\maketitle
	
\section{Introduction}
\label{sec:intro}
	
The hierarchy between the electroweak and Planck scales remains a central motivation for physics beyond the Standard Model. Braneworld scenarios provide a possible mechanism to address this problem, generically predicting modifications of gravity at short distances that manifest as departures from the Newtonian inverse-square law. A key feature of these scenarios is that distinct Kaluza--Klein (KK) spectra map to distinct distance dependences of the short-range potential correction, parameterized by a dimensionless term $\Delta(r)$ such that $U(r)=U_{\rm N}(r)\,[1+\Delta(r)]$. In this work, we focus on three representative KK spectral classes that capture the main phenomenological possibilities at laboratory scales:
    (i) the gapless warped continuum of Randall--Sundrum II (RSII)~\cite{Randall:1999vf,Garriga:1999yh,Callin:2004py}, which produces a power-law tail;
	(ii) discrete KK towers from compact extra dimensions of Arkani-Hamed--Dimopoulos--Dvali (ADD) type~\cite{Antoniadis:1998ig,ArkaniHamed:1998rs}, which yield a compactification-controlled enhancement;
	and (iii) the warped thick-brane continuum with a finite mass gap $\lambda_{\rm gap}=m_{\rm gap}^{-1}$.
	The latter class is typically realized by P\"oschl--Teller (PT)-type potentials across various thick-brane constructions, including standard Einstein--scalar, Weyl-integrable,  and mimetic frameworks~\cite{Gremm:2000dj,BarbosaCendejas:2007MassGap,BarbosaCendejas:2008WeylMassGap,Sui:2021mimetic}, generating a Yukawa-like exponential suppression.
	
	Precision short-range tests of Newtonian gravity, most notably torsion-balance experiments, have placed stringent constraints on such macroscopic deviations~\cite{Adelberger:2009zz,Kapner:2006si,Tan:2016,Lee:2020,Tan:2020}.
	However, pushing sensitivities below tens of microns with traditional macroscopic experiments is increasingly hindered by electromagnetic backgrounds and geometric systematics, including Casimir and patch-potential effects~\cite{Klimchitskaya:2009RMP,Behunin:2014Patch}.
	This limitation strongly motivates complementary probes that access the gravitational interaction through different observables.
	
Quantum-gravity-induced entanglement of masses (QGEM), originally proposed as an entanglement witness for the quantized nature of the gravitational field~\cite{Bose:2017SpinWitness,MarlettoVedral:2017BMV,MarlettoVedral:2025RMP,Bose:2025Interfaces}, offers such an alternative. While there is an ongoing theoretical debate regarding the interpretation of QGEM in the presence of semiclassical or hybrid models~\cite{KafriTaylorMilburn:2014ClassicalChannel,TilloyDiosi:2017LeastDecoherence,AnastopoulosHu:2022Entropy,AzizHowl:2025Nature,Chevalier:2020UnknownInteractions,MarlettoVedral:2025LocalMeans,MarlettoOppenheimVedralWilson:2025ClassicalCannot,Diosi:2025NoClassicalGravityEntangles}, the low-energy dynamics of the protocol are captured by branch-dependent phase accumulation in a matter-wave interferometric setting. Crucially, the entangling phase is extraordinarily sensitive to the exact shape of the central potential. In the small-splitting regime $\Delta x\ll d$ relevant to near-term interferometry (where $\Delta x$ is the wavepacket splitting within each mass and $d$ is the mean separation between the masses), the phase effectively probes not only the absolute size of the short-range correction at $r\simeq d$, but also how rapidly it varies with separation (its local slope).
	
Because QGEM is inherently suited to detecting interactions with steep distance dependence, it serves as a powerful precision probe for non-Newtonian interactions. Entanglement-based tomography proposals track tiny interaction-induced phases to constrain Yukawa-type macroscopic forces, while explicitly accounting for Coulomb and Casimir--Polder backgrounds~\cite{Marshman:2022Tomography}. Complementary entanglement-witness analyses have been developed for Yukawa potentials mediated by axion-like particles, including realistic decoherence budgeting~\cite{CarmonaRufo:2025ALPWitness}. QGEM-type entanglement signatures have also been explored in extra-dimensional gravity, emphasizing the roles of massless versus massive gravitons in warped geometries and RSII-type settings~\cite{Elahi:2023ProbingMassive,Feng:2024QGEMExtraDim}. 

Driven by rapid experimental progress on QGEM-type platforms, this work systematically evaluates and distinguishes the aforementioned three KK spectral classes using entanglement observables. We complement previous studies by comparing these models within a common comparative framework: the ADD and RSII reference values are tied to representative short-range gravity bounds, while the PT-type mass-gap benchmark $\lambda_{\rm gap}=94\,\mu{\rm m}$ follows Ref.~\cite{Sui:2021mimetic}.
		
To evaluate the potential experimental reach, we consider both conservative and optimistic near-term configurations for the mass, superposition size, and coherence time $(m, \Delta x, t)$. These choices are informed by recent advancements in background mitigation and parallel interferometric geometries~\cite{vanDeKamp:2020PRA,Schut:2023PRR,Schut:2024Micrometer}. Under these configurations, we evaluate the fractional entangling-phase shift, concurrence, and normalized phase-response profiles as functions of the separation $d$. Our results reveal a consistent signal hierarchy among the three benchmarks, with the ADD model yielding the most pronounced effects. 
	
The paper is organized as follows.
Section~\ref{sec:SR-modifications} summarizes the three classes of short-range modifications and the benchmark parameter choices adopted here.
Section~\ref{sec:QGEM} reviews the QGEM protocol and derives model-independent expressions for the entangling phase and concurrence.
Section~\ref{sec:Results} evaluates the QGEM signals under conservative and optimistic experimental parameters, complemented by an analysis of normalized phase-response profiles for distance-scan model discrimination.
We conclude in Sec.~\ref{sec:Summary} with a summary and outlook.

\section{Extra-dimensional corrections to Newtonian gravity at short range}
\label{sec:SR-modifications}
	
We parameterize the static interaction between two point masses $m_1,m_2$ localized on the brane as
\begin{equation}
		U(r)\;=\; U_N(r)\,\bigl[1+\Delta(r)\bigr],
		\label{eq:U-master}
\end{equation}
where $U_N(r)\equiv -G_N m_1 m_2/r$ is the Newtonian potential, and $\Delta(r)$ denotes the correction induced by KK gravitons. We summarize below three representative extra-dimensional spectra and the corresponding
	short-range modifications of Newtonian gravity, together with the benchmark parameter choices adopted in this work.
	
Fig.~\ref{fig:SecII_spectra_potentials} provides a schematic overview of the tensor-mode potentials and KK spectra for the three benchmark scenarios considered in this work: RSII (gapless KK continuum), ADD (discrete KK tower), and thick branes with a mass gap (gapped KK continuum).

\subsection{RSII: gapless KK continuum}
\label{subsec:RSII}
	
\begin{figure}[tbp]
\centering
	\subfloat[]{\includegraphics[width=7cm,height=4.2cm]{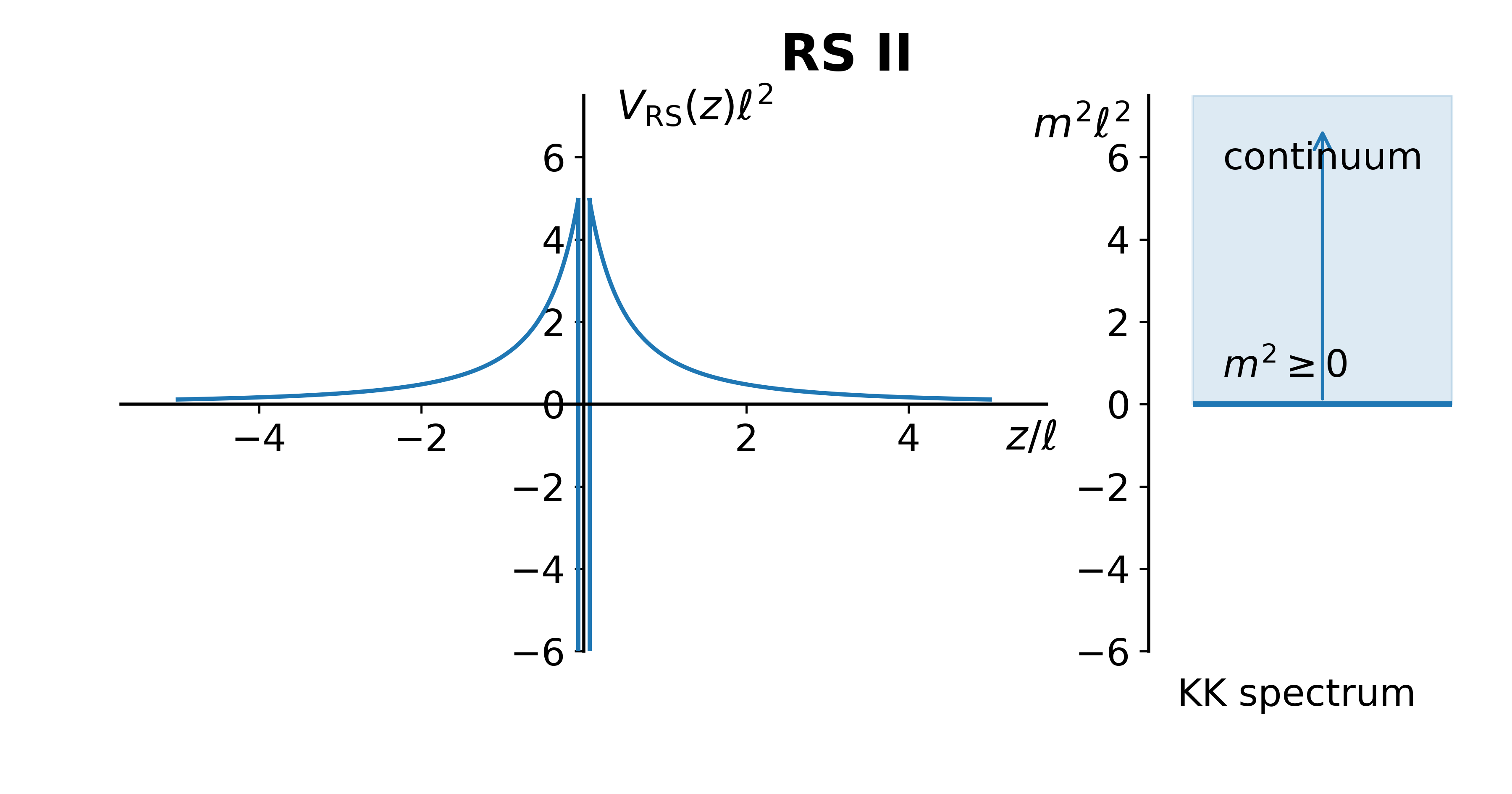}}\\
	\subfloat[]{\includegraphics[width=7cm,height=4.2cm]{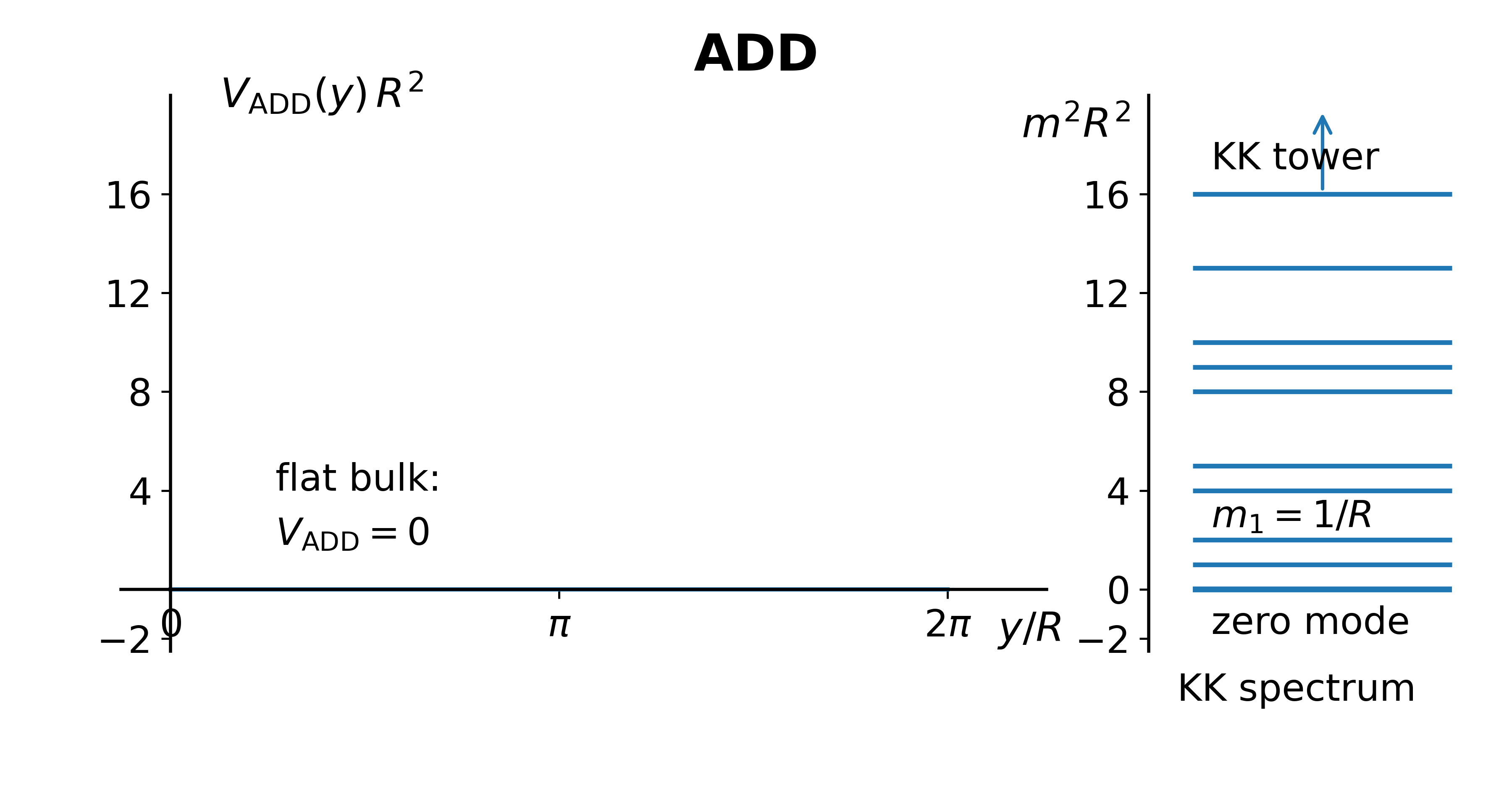}}\\
	\subfloat[]{\includegraphics[width=7cm,height=4.2cm]{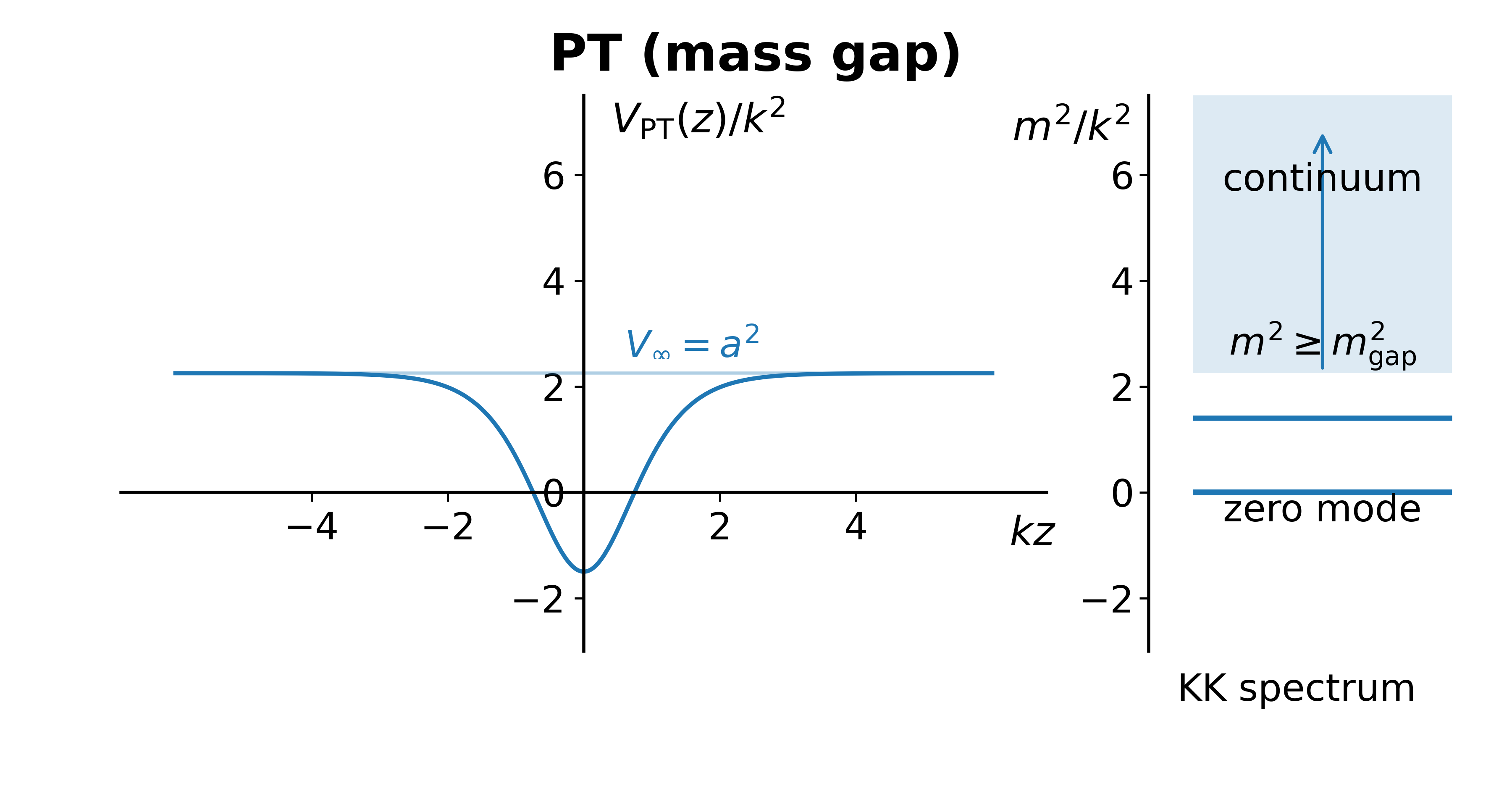}}
\caption{Schematic tensor-mode potential (left of each panel) and the corresponding KK spectrum (right of each panel) for the three benchmark scenarios considered in Sec.~II. (a) RSII: the tensor sector can be cast into a Schr\"odinger-like problem with a volcano potential approaching zero at large $|z|$; the brane-localized $\delta$-function well is indicated schematically, implying a localized zero mode plus a gapless continuum $m^2\ge 0$. (b) ADD ($\delta=2$): a flat bulk with compact extra dimensions, where the KK masses are quantized as $m_{\mathbf{n}}=|\mathbf{n}|/R$, yielding a discrete tower; there is no warp-induced localization potential ($V_{\rm ADD}=0$). (c) Thick branes with a mass gap: illustrated by the modified PT benchmark in Eq.~(\ref{eq:VPT}), which approaches a positive plateau $V_\infty=m_{\rm gap}^2=a^2k^2$, so that the continuum starts at $m^2\ge m_{\rm gap}^2$.}
\label{fig:SecII_spectra_potentials}
\end{figure}

The RSII model features a single positive-tension brane embedded in a five-dimensional AdS bulk with curvature radius $\ell$ (equivalently $k_{\rm RS}=\ell^{-1}$)~\cite{Randall:1999vf}. A convenient coordinate choice yields the warped background
\begin{equation}
	ds^{2}=e^{-2|y|/\ell}\,\eta_{\mu\nu}dx^{\mu}dx^{\nu}+dy^{2},
\label{eq:RSII_metric}
\end{equation}
where $y$ denotes the coordinate of extra dimension. After the standard KK decomposition and a change to the conformal coordinate $z$, linearized transverse-traceless tensor perturbations reduce to a one-dimensional Schr\"odinger-like problem,
\begin{equation}
	\left[-\partial_{z}^{2}+V_{\rm RS}(z)\right]\psi_{m}(z)=m^{2}\psi_{m}(z),
\label{eq:RSII_schro}
\end{equation}
with the ``volcano''  potential
\begin{equation}
	V_{\rm RS}(z)=\frac{15}{4(|z|+\ell)^{2}}-\frac{3}{\ell}\,\delta(z),
\label{eq:RSII_potential}
\end{equation}
see, e.g., Refs.~\cite{Garriga:1999yh,Callin:2004py}.
The tensor spectrum consists of a localized massless graviton with $\psi_{0}(z)\propto (|z|+\ell)^{-3/2}$ and a \emph{gapless} continuum of KK modes starting at $m=0$~\cite{Garriga:1999yh,Callin:2004py}. See Fig.~\ref{fig:SecII_spectra_potentials}(a) for a schematic of the volcano potential and the gapless KK continuum.
	
The resulting correction to the Newtonian potential can be expressed as \cite{Callin:2004py}
\begin{equation}
	\Delta_{\rm RS}(r)=
		\frac{8}{3\pi^2}\int_{0}^{\infty}\!dn\,
		\frac{e^{-n r/\ell}}{n\bigl[J_1^2(n)+Y_1^2(n)\bigr]},
		\label{eq:DeltaRS_integral}
\end{equation}
where $n$ is a dimensionless KK spectral parameter, related to the continuum KK mass by $m=n/\ell$;  $J_1$ and $Y_1$ denote the Bessel functions of the first and second kind of order one. Equation~\eqref{eq:DeltaRS_integral} is a spectral representation of the KK correction in which each KK mass scale contributes with a Yukawa factor $e^{-mr}$. Because the RSII KK spectrum is \emph{gapless} (starting at $m=0$), the small-$m$ region controls the long-distance behavior and yields a power-law tail rather than a single Yukawa fall-off. It therefore reduces at large separations to the well-known power-law behavior~\cite{Garriga:1999yh,Callin:2004py}
\begin{equation}
	\Delta_{\rm RS}(r)\;=\;\frac{2\ell^2}{3r^2}
		+\mathcal{O}\!\left(\frac{\ell^4}{r^4}\ln\frac{r}{\ell}\right),
		\qquad (r\gg \ell).
\label{eq:DeltaRS_asympt}
\end{equation}
	
At short distances $r\ll \ell$, the potential crosses over to the five-dimensional regime $U(r)\propto 1/r^{2}$, so the KK contribution is no longer a small perturbation to the Newtonian form (see Refs.~\cite{Garriga:1999yh,Callin:2004py}). All numerical results below use the full representation in Eq.~\eqref{eq:DeltaRS_integral} rather than asymptotic limits.
	
Submillimeter tests of the inverse-square law directly constrain the AdS curvature radius. Torsion-balance measurements reaching separations down to $r\simeq 52\,\mu{\rm m}$ \cite{Kapner:2006si,Lee:2020} impose the bound on the AdS curvature radius as
\begin{equation}
	\ell \;\le\; 52\,\mu{\rm m}.
	\label{eq:RSII_bound}
\end{equation}
We adopt this as the representative short-range benchmark value for the RSII curvature scale used below.
	
This mass spectrum structure extends broadly to many thick-brane models, which can be viewed as smooth generalizations of the RSII scenario. These models feature a similar tensor spectrum consisting of a normalizable zero mode and a gapless KK continuum starting at $m=0$. In such cases, the KK correction produces a power-law tail at large distances, making the resulting short-range potential qualitatively identical to the RSII result up to model-dependent $\mathcal{O}(1)$ coefficients, as emphasized in Ref.~\cite{Csaki:2000fc}.

\subsection{ADD: discrete KK tower}
\label{subsec:ADD}
	
In ADD models, gravity propagates in $4+\delta$ dimensions with $\delta$ compact flat extra dimensions of radius $R$, while Standard Model fields are confined to the brane \cite{ArkaniHamed:1998rs,Antoniadis:1998ig}. The higher-dimensional background is taken to be flat, as described by the metric
\begin{equation}
	ds^{2}=\eta_{\mu\nu}dx^{\mu}dx^{\nu}+\sum_{i=1}^{\delta}dy_{i}^{2},
		\quad y_{i}\simeq y_{i}+2\pi R.
\label{eq:ADD_metric}
\end{equation}
Thus, the KK gravitons correspond to Fourier modes along the compact directions.

Compactification gives rise to a discrete KK tower with masses $m_{\vec n}=|\vec n|/R$, $\vec n\in\mathbb{Z}^\delta$. A convenient representation of the KK correction is an explicit sum over Yukawa exchanges of the KK tower (see, e.g., Refs.~\cite{ArkaniHamed:1998rs,Antoniadis:1998ig,Navas:2024RPP}):
\begin{equation}
	\Delta_{\rm ADD}(r)=
		\frac{4}{3}\sum_{\vec n\neq \vec 0}
		\exp\!\left(-\frac{|\vec n|\,r}{R}\right),
		\label{eq:DeltaADD_sum}
\end{equation}
where the factor $4/3$ accounts for the polarization structure of massive spin-2 exchange.
	
The compactification radius implied by a given fundamental scale depends strongly on the number $\delta$ of flat extra dimensions. The observed effective four-dimensional Planck scale is related to the fundamental $(4+\delta)$-dimensional Planck scale $M_{4+\delta}$ by
\begin{equation}
	M_{\rm Pl}^{2}\simeq (2\pi R)^{\delta}\,M_{4+\delta}^{2+\delta},
\label{eq:ADD_planck_relation}
\end{equation}
up to order-unity factors that depend on the compactification geometry (see, e.g., Ref.~\cite{Navas:2024RPP}). For $\delta=1$, reproducing the observed four-dimensional Planck scale requires an astronomical-sized radius, excluded by macroscopic tests of gravity. For $\delta\ge 3$, the corresponding radii lie far below the tens-of-microns window, so the KK sum in Eq.~\eqref{eq:DeltaADD_sum} is exponentially suppressed at laboratory separations. Consequently, short-range inverse-square-law experiments most directly constrain the case $\delta=2$. Current torsion-balance experiments constrain the compactification radius to
\begin{equation}
	R \le 30\,\mu{\rm m},
\label{eq:ADD_bound}
\end{equation}
based on the absence of deviations from Newtonian gravity at submillimeter scales~\cite{Kapner:2006si,Tan:2016,Lee:2020,Tan:2020}. This bound sets the benchmark ADD reference value used in the following analysis.
	
Equation~\eqref{eq:DeltaADD_sum} interpolates between two physically distinct regimes. At separations $r\ll R$, the KK spectrum is effectively quasi-continuous and the gravitational field lines explore the full $(4+\delta)$-dimensional bulk, so the potential scales as $U(r)\propto r^{-(1+\delta)}$. For the laboratory-relevant case $\delta=2$, this corresponds to the characteristic six-dimensional behavior $U(r)\propto 1/r^{3}$, and hence
\begin{equation}
	\Delta_{\rm ADD}(r)\propto (R/r)^{2},	
\end{equation}
up to geometry-dependent ${\cal O}(1)$ factors. At $r\gg R$, the correction is dominated by the lightest KK levels with $m\sim 1/R$ and becomes Yukawa-like, 
\begin{equation}
	\Delta_{\rm ADD}(r)\propto e^{-r/R},
\end{equation}	
with heavier shells providing subleading contributions.
	
For the benchmark baselines considered in this work ($d\simeq 40$--$80\,\mu{\rm m}$), representative short-range bounds still allow $R$ in the tens-of-microns range for $\delta=2$, so one often probes the crossover window $d/R=\mathcal{O}(1)$. In this regime it is essential to evaluate Eq.~\eqref{eq:DeltaADD_sum} as an explicit KK sum rather than relying on either limiting approximation.

\subsection{Gapped KK continuum from thick branes}
\label{subsec:MassGap}
	
A qualitatively distinct KK spectrum can arise in smooth warped thick-brane backgrounds of the generic form
\begin{equation}
	ds^{2}=e^{2A(y)}\,\eta_{\mu\nu}dx^{\mu}dx^{\nu}+dy^{2}.
\label{eq:thickbrane_metric}
\end{equation}
For transverse-traceless tensor perturbations, after the standard KK decomposition  and a change to the conformal coordinate $z$ defined by $dz=e^{-A(y)}dy$, the KK profiles obey a one-dimensional Schr\"odinger equation,
\begin{align}
	\left[-\partial_{z}^{2}+V_T(z)\right]\psi_{m}(z) &= m^{2}\psi_{m}(z), \\
		V_T(z) &= \frac{3}{2}A''(z)+\frac{9}{4}A'(z)^{2},
\label{eq:thickbrane_tensor_schro}
\end{align}
where the prime denotes derivatives with respect to $z$. If $V_T(z\to\infty)\to m_{\rm gap}^{2}>0$, the spectrum contains a localized massless graviton and a continuous KK sector separated from zero by a finite mass gap $m_{\rm gap}$ (a gapped continuum)~\cite{Nam:1999MassGapKK}. In many analytically tractable realizations, the potential is well approximated by a modified PT form, providing a convenient benchmark for gapped continuum.
	
PT-type potentials and gapped continuum arise broadly across thick-brane realizations, appearing in standard five-dimensional Einstein--scalar domain walls ~\cite{Gremm:2000dj,BarbosaCendejas:2007MassGap} as well as extended gravitational frameworks such as Weyl-integrable geometry~\cite{BarbosaCendejas:2008WeylMassGap} and mimetic thick-brane constructions~\cite{Sui:2021mimetic}. This motivates treating the gapped continuum as a representative spectral class characterized phenomenologically by the single length scale $\lambda_{\rm gap}=m_{\rm gap}^{-1}$.
	
For the PT-type gapped thick-brane class considered here, a convenient representative parametrization is obtained from the warp factor
\begin{equation}
	e^{A(z)}=\sech(kz),
\label{eq:warp_PT_mimetic}
\end{equation}
which yields the modified PT tensor potential, as realized for example in Ref.~\cite{Sui:2021mimetic},
\begin{equation}
	V_{\rm PT}(z)=a^{2}k^{2}-a(a+1)k^{2}\sech^{2}(kz),
\label{eq:VPT}
\end{equation}
with $a=\frac{3}{2}$. In the limit $z\to\pm\infty$, $V_{\rm PT}\to a^{2}k^{2}$, implying that the mass continuum begins at the threshold 
\begin{equation}
	m_{\rm gap}=a\,k=\frac{3}{2}k,
\end{equation}
Correspondingly, the characteristic length scale associated with this mass gap can be defined as
\begin{equation}
	\lambda_{\rm gap}\equiv m_{\rm gap}^{-1}=\frac{2}{3k}.
\label{eq:mgap_lambdagap}
\end{equation}

Besides the localized zero mode, this PT potential also contains one odd-parity bound state with mass $m_{1}=\sqrt{2}\,k$. Because $\psi_{1}(0)=0$, that odd state does not contribute to the gravitational potential between sources located on the brane, and the KK correction is controlled entirely by the continuum modes with $m\ge m_{\rm gap}$.
	
For this case, the correction induced by the gapped continuum is given by~\cite{Sui:2021mimetic}
\begin{equation}
	\Delta_W(r)
	=
	\left(\int_{-\infty}^{+\infty}dz\,e^{3A(z)}\right)
	\int_{m_{\rm gap}}^{\infty}dm\,e^{-mr}\,|\psi_m(0)|^{2}.
\label{eq:DeltaW_general}
\end{equation}
Using Eq.~\eqref{eq:warp_PT_mimetic}, one has
\begin{equation}
	\int_{-\infty}^{+\infty}dz\,e^{3A(z)}
	=
	\int_{-\infty}^{+\infty}dz\,\sech^{3}(kz)
	=
	\frac{\pi}{2k},
\label{eq:sech3_integral}
\end{equation}
so the exact continuum representation can be written as
\begin{equation}
	\!\!\!\Delta_W(r)
	\!=\!
	\frac{\pi}{2k}
	\int_{m_{\rm gap}}^{\infty}dm \, e^{-mr}
	\left|
	\frac{\Gamma(1-\sigma)}
	{\Gamma\!\left(-\frac14\!-\!\frac{\sigma}{2}\right)
		\Gamma\!\left(\frac74\!-\!\frac{\sigma}{2}\right)}
	\right|^{2}\!\!,
\label{eq:DeltaW_spectral}
\end{equation}
where $\sigma(m)\equiv\sqrt{\frac94-\frac{m^{2}}{k^{2}}}$ is defined via analytic continuation for $m\ge m_{\rm gap}$ to ensure that the integrand remains well-defined throughout the continuum.  
	
Equation~\eqref{eq:DeltaW_spectral} makes the physical origin of the short-range behavior transparent: because the KK continuum starts at the finite threshold $m_{\rm gap}$, the correction is Yukawa suppressed at large separations,
\begin{equation}
	\Delta_W(r)\propto \frac{e^{-m_{\rm gap}r}}{kr}=\frac{e^{-r/\lambda_{\rm gap}}}{kr},
		\qquad (r\gg \lambda_{\rm gap}),
\end{equation}
in agreement with the large-distance behavior found in Ref.~\cite{Sui:2021mimetic}.

It is convenient to factor out the exponential threshold suppression and define a slowly varying effective strength,
\begin{equation}
	\Delta_W(r)=\alpha_W(r)\,e^{-m_{\rm gap}r},
\label{eq:alphaW_def}
\end{equation}
where $\alpha_W(r)\equiv e^{m_{\rm gap}r}\Delta_W(r)$. Therefore, unlike a pure Yukawa correction with constant prefactor, the PT continuum retains a mild residual $r$-dependence through $\alpha_W(r)$.
	This residual running will later modify the amplification factor $\mathcal{A}(d)$ beyond the pure-Yukawa baseline.
	
Reference~\cite{Sui:2021mimetic} further showed, by comparing the PT-type correction with the latest short-range inverse-square-law tests, that the mimetic thick-brane parameter is constrained approximately by $k \gtrsim 1.4\times 10^{-3}\ {\rm eV}$. This translates into a benchmark mass-gap scale of
\begin{equation}
	\lambda_{\rm gap}\lesssim 94\,\mu{\rm m}.
\label{eq:lambda_gap_benchmark}
\end{equation}
In the subsequent numerical analysis,  we adopt this benchmark and evaluate the gapped-continuum contribution using the exact spectral form in Eq.~\eqref{eq:DeltaW_spectral}.

\section{QGEM protocol and entangling-phase observables}
\label{sec:QGEM}
	
Quantum-gravity-induced entanglement of two mesoscopic masses provides a phase-based witness of gravitational interactions in the weak-field, nonrelativistic regime~\cite{Bose:2017SpinWitness,MarlettoVedral:2017BMV}. In this section, we summarize the protocol and collect the model-independent expressions that map a
	central potential $U(r)$ into an entangling phase $\Phi(d)$ and the concurrence $C(d,t)$.
	
\subsection{QGEM setup and phase accumulation}
\label{subsec:QGEM_setup}
	
\begin{figure}[tbp]
\centering
	 \includegraphics[width=\linewidth]{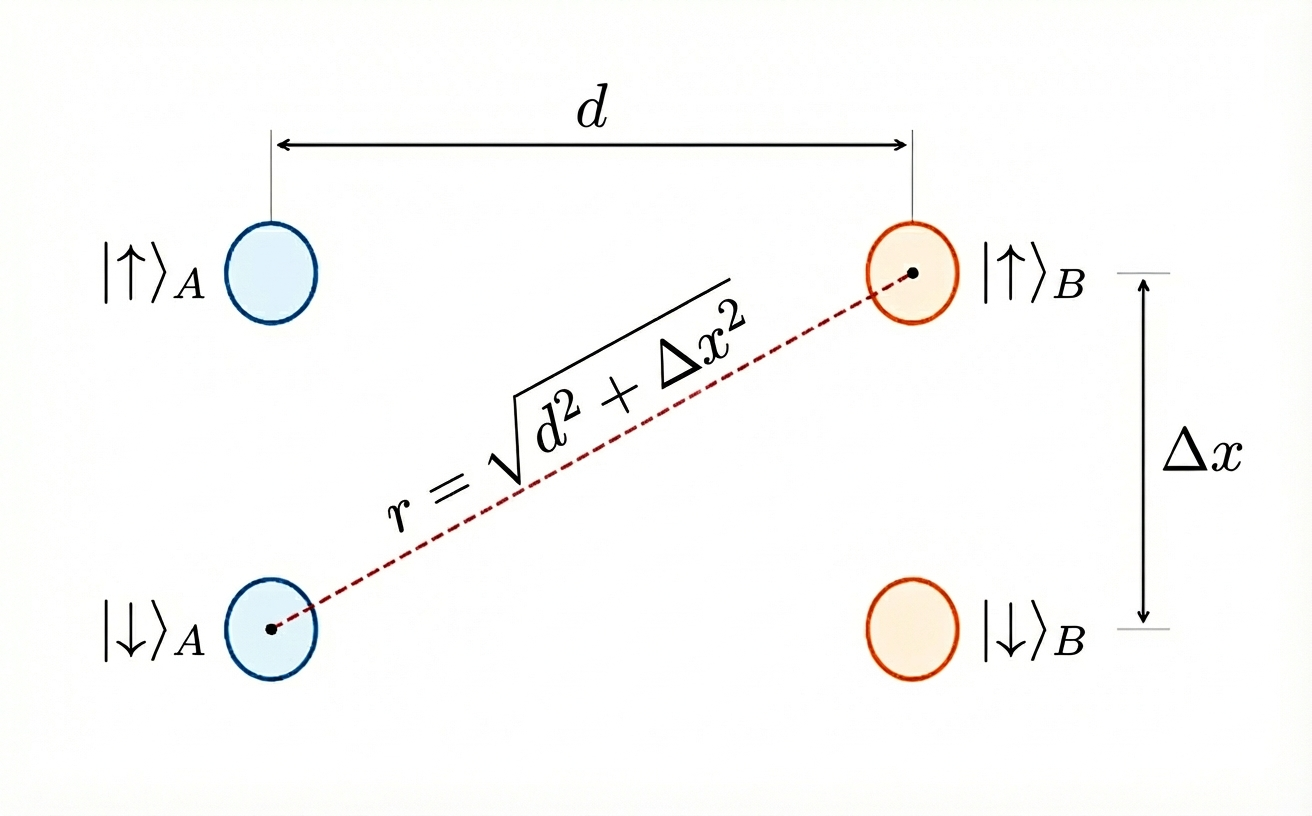}
\caption{Schematic of the parallel QGEM configuration. Two test masses, $A$ and $B$, are prepared in spatial superpositions of localized wavepackets with a transverse splitting $\Delta x$. The centers of the two systems are separated by a baseline distance $d$. Gravitational interaction over a time $t$ induces branch-dependent phases $\phi_{ss'}$ determined by the respective separations $r_{ss'}$ through the potential $U(r)$.}
\label{fig:qgem_diagram}
\end{figure}

We consider two identical masses $m_1=m_2=m$. Each of the two masses, labeled $A$ and $B$, is prepared in a superposition of two localized wave packets separated by a distance
	$\Delta x$ and correlated with an internal two-level degree of freedom
	$\{\ket{\uparrow},\ket{\downarrow}\}$.
	The mean separation between the two systems is denoted by $d$ (parallel configuration),
	as illustrated in Fig.~\ref{fig:qgem_diagram}.
	After the initial beam splitters, the joint state is
\begin{equation}
	\ket{\psi_0}\;=\;
	\frac{1}{2}\bigl(\ket{\uparrow}_A+\ket{\downarrow}_A\bigr)
		\bigl(\ket{\uparrow}_B+\ket{\downarrow}_B\bigr).
\label{eq:psi0_QGEM}
\end{equation}
During an interaction time $t$, each branch $(s,s')\in\{\uparrow,\downarrow\}^2$ accumulates a phase
\begin{equation}
		\phi_{ss'}(d)
			= \frac{t}{\hbar}\,U\!\left(r_{ss'}\right), 
\label{eq:branch_phase_def}
\end{equation}
where $r_{ss'}$ denotes the branch-dependent separation.
	
After the interaction time $t$, the branch-dependent phases imprint on the joint internal state,
\begin{equation}
	\ket{\psi(t)}\;=\;
		\frac{1}{2}\sum_{s,s'\in\{\uparrow,\downarrow\}}
		e^{-i\phi_{ss'}(d)}\,
		\ket{s}_A\ket{s'}_B,
\label{eq:psi_t_phases}
\end{equation}
where  the global phase is physically irrelevant and local (single-particle) phases can be absorbed into local operations on $A$ and $B$.
	
For the geometry in Fig.~\ref{fig:qgem_diagram}, the separations are
\begin{equation}
	r_{\uparrow\uparrow}=r_{\downarrow\downarrow}=d,
		\qquad
	r_{\uparrow\downarrow}=r_{\downarrow\uparrow}=\sqrt{d^{2}+\Delta x^{2}}.
\label{eq:branch_distances}
\end{equation}
	
The entanglement is controlled by the entangling phase~\cite{Bose:2017SpinWitness,MarlettoVedral:2017BMV}
\begin{equation}
	\Phi(d)
		\;\equiv\;
		\phi_{\uparrow\downarrow}(d)+\phi_{\downarrow\uparrow}(d)
		-\phi_{\uparrow\uparrow}(d)-\phi_{\downarrow\downarrow}(d).
\label{eq:Phi_def_general}
\end{equation}
Using Eq.~\eqref{eq:branch_distances}, this reduces to
\begin{equation}
	\Phi(d)
		\;=\;
		\frac{2t}{\hbar}\Bigl[
		U\!\left(\sqrt{d^{2}+\Delta x^{2}}\right)-U(d)
		\Bigr].
\label{eq:Phi_exact}
\end{equation}
	
\subsection{Entangling phase from a general central potential}
\label{subsec:Phi_general_potential}
	
Equation~\eqref{eq:Phi_exact} applies to any central potential $U(r)$. In the experimentally relevant regime $\Delta x\ll d$, it is useful to introduce the small geometric offset
\begin{equation}
		\delta r \;\equiv\; \sqrt{d^{2}+\Delta x^{2}}-d
		\;=\;\frac{\Delta x^{2}}{2d}+\mathcal{O}\!\left(\frac{\Delta x^{4}}{d^{3}}\right).
		\label{eq:deltar_def}
\end{equation}
Expanding $U(r)$ about $r=d$ and retaining only the leading-order term in $\delta r$, we obtain
\begin{align}
	\Phi(d)
	&= \frac{2t}{\hbar}\Bigl[U(d+\delta r)-U(d)\Bigr]\simeq \frac{2t}{\hbar}\,\delta r\,U'(d)\nonumber \\
	&\simeq \frac{t}{\hbar}\,\frac{\Delta x^{2}}{d}\,U'(d),
			\qquad (\Delta x \ll d),
\label{eq:Phi_smallDx}
\end{align}
where $U'(d)\equiv \left.\frac{dU}{dr}\right|_{r=d}$. For the Newtonian potential, the derivative evaluates to $U_N'(d)={G_N m^2}/{d^2}$, and the Newtonian entangling phase reduces to
\begin{equation}
	 \Phi_N(d)
	 \simeq \frac{t}{\hbar}\,\frac{\Delta x^{2}}{d}\,U_N'(d)=
	\frac{G_N m^2 t}{\hbar}\,\frac{\Delta x^{2}}{d^{3}}.
\label{eq:Phi_Newton}
\end{equation}
Consequently, the ratio of the entangling phases is uniquely governed by the functional form of the gravitational potentials, i.e.,
\begin{equation}
	\frac{\Phi(d)}{\Phi_N(d)}
	\;\simeq\;
	\frac{U'(d)}{U_N'(d)}.
\label{eq:Phi_ratio_smallDx}
\end{equation}

To isolate the non-Newtonian signatures, we define the non-Newtonian entangling-phase shift as $\delta\Phi(d) \equiv \Phi(d) - \Phi_N(d)$ and parameterize the deviation through the fractional entangling-phase shift $\delta\Phi(d)/\Phi_N(d)$. From Eq.~\eqref{eq:Phi_ratio_smallDx}, this fractional phase shift can be expressed as
\begin{equation}
	\frac{\delta\Phi(d)}{\Phi_N(d)}
	\;\simeq\;
	\frac{U'(d)}{U_N'(d)}-1.
\label{eq:Relat_phi_ratio_smallDx}
\end{equation}
Differentiating the potential in Eq.~\eqref{eq:U-master} yields
\begin{equation}
		\frac{U'(r)}{U_N'(r)}
		\;=\;
		1+\Xi(r),
		\label{eq:Uprime_ratio}
\end{equation}	
where $\Xi(r)\equiv \Delta(r)-r\,\Delta'(r)$. We refer to $\Xi(d)$ as the phase-response profile, since it encodes the leading distance-dependent response of the QGEM entangling phase to the non-Newtonian correction. Thus, the fractional entangling-phase shift reads
\begin{equation}
	\frac{\delta\Phi(d)}{\Phi_N(d)}\;\simeq\;\Xi(d).
	\label{eq:Xi_def}
\end{equation} 

While short-range modifications are often characterized at the level of the potential itself via $\Delta(d)$, the phase-response profile effectively reweights this correction. To quantify how this derivative-dependent response modulates the signal, we define the amplification factor
\begin{equation}
	\mathcal{A}(d)\;\equiv\;\frac{\Xi(d)}{\Delta(d)}
		\;=\;
		1-\frac{d\,\Delta'(d)}{\Delta(d)}.
\label{eq:Amplification_def}
\end{equation}
For a pure Yukawa form $\Delta(d)=\alpha\,e^{-d/\lambda}$, one has $\mathcal{A}(d)=1+d/\lambda$, illustrating the growing sensitivity of $\Xi(d)$ at separations larger than the range. For power-law tails $\Delta\propto d^{-p}$, $\mathcal{A}=1+p$ becomes constant. In gapped continuum, the additional running of the effective strength $\alpha_{\rm W}(d)$ further modifies $\mathcal{A}(d)$ beyond the pure-Yukawa baseline.

\begin{figure}[tbp]
\centering
\includegraphics[width=0.95\linewidth]{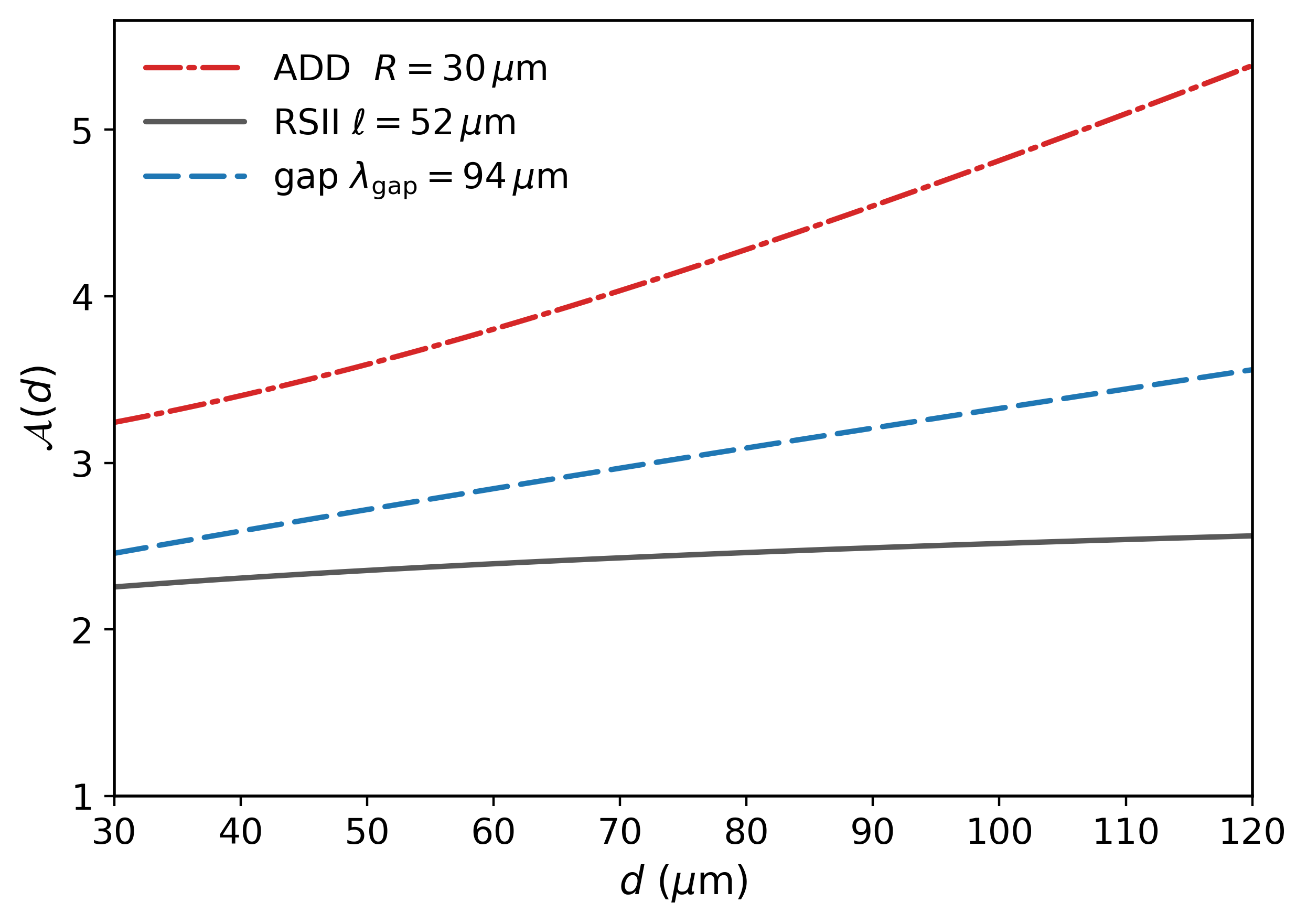}
\caption{Amplification factor $\mathcal{A}(d)$. Curves are shown for the benchmark parameter choices used below: ADD ($\delta=2$, $R=30\,\mu{\rm m}$), RSII ($\ell=52\,\mu{\rm m}$), and the gapped continuum ($\lambda_{\rm gap}=94\,\mu{\rm m}$).}
\label{fig:amplification_factor}
\end{figure}

Figure~\ref{fig:amplification_factor} visualizes $\mathcal{A}(d)$ for the three spectral classes at representative benchmark points. Across the displayed range, all three curves increase monotonically with $d$ and follow the hierarchy ADD $>$ gapped $>$ RSII. This ordering indicates that the gapped continuum receives a more pronounced derivative boost than the RSII model.
	
By connecting the potential-level correction $\Delta(d)$ to the phase-response profile $\Xi(d)$, the amplification factor $\mathcal{A}(d)$ provides a compact measure of derivative amplification.  However, it is best viewed as a diagnostic of this enhancement rather than a standalone reach metric, because the absolute phase signal remains governed by the product $\Xi(d)=\mathcal{A}(d)\,\Delta(d)$.
	
\subsection{Concurrence as a reach metric}
\label{subsec:Concurrence_reach}
	
In the ideal protocol, the reduced state of the two internal qubits after recombination is locally equivalent to a controlled-phase entangled state whose entanglement is governed by $\Phi(d)$. A convenient entanglement monotone is the concurrence~\cite{Elahi:2023ProbingMassive,Feng:2024QGEMExtraDim},
	which for the present equal-amplitude four-branch superposition takes the form
\begin{equation}
	C(d,t)=
		\left|\sin\!\left[\frac{\Phi(d,t)}{2}\right]\right|.
\label{eq:Concurrence_from_Phi}
\end{equation}
In the small-phase regime $\Phi(d,t)\ll 1$, one has
\begin{equation}
	C(d,t)\simeq \frac{|\Phi(d,t)|}{2}.
\label{eq:Concurrence_smallPhi}
\end{equation}
To characterize the experimental reach, we introduce a phase-resolution benchmark $\Phi_{\rm det}$ and define a corresponding detection threshold as
\begin{equation}
	C_{\rm det}\equiv\left|\sin\!\left(\frac{\Phi_{\rm det}}{2}\right)\right|.
\label{eq:Cdet_def_results}
\end{equation}
The specific benchmark values for $(m,\Delta x,t)$ and $\Phi_{\rm det}$ are provided alongside the numerical results in the following section.

\section{QGEM signals for benchmark parameter choices}
\label{sec:Results}
	
This section presents numerical results for the entangling phase and concurrence induced by the modified potential in Eq.~\eqref{eq:U-master}. Following the general QGEM framework derived in last section, we evaluate the fractional entangling-phase shift $\delta\Phi(d)/\Phi_N(d)$ and the concurrence $C(d)$ as functions of the separation $d$ for the three spectral classes introduced in Sec.~\ref{sec:SR-modifications}. To quantify the experimental reach, we consider two representative sets of QGEM parameters $(m,\Delta x,t)$. These include a conservative configuration with robust margins against non-gravitational systematics as well as a more optimistic near-term setting that accounts for plausible improvements in shielding and coherence times~\cite{vanDeKamp:2020PRA,Schut:2023PRR,Schut:2024Micrometer}. The first two subsections present the absolute phase and concurrence signals for these two representative experimental settings. We then turn to a complementary analysis based on the normalized phase-response profile, which is less sensitive to overall normalization and more directly probes model discrimination.

\subsection{Conservative experimental parameters}
\label{subsec:cons_bench}
	
We first adopt conservative experimental parameters,
\begin{equation}
		m=1\times10^{-14}\,{\rm kg},\quad
			\Delta x=10\,\mu{\rm m},\quad
			t=0.10\,{\rm s},
\label{eq:bench_cons}
\end{equation}
which are chosen to stay comfortably within the ranges discussed in Refs.~\cite{vanDeKamp:2020PRA,Schut:2023PRR,Schut:2024Micrometer} while ensuring sensitivity to submillimeter modifications of gravity.

\begin{figure}[tbp]
\centering
\includegraphics[width=1\linewidth]{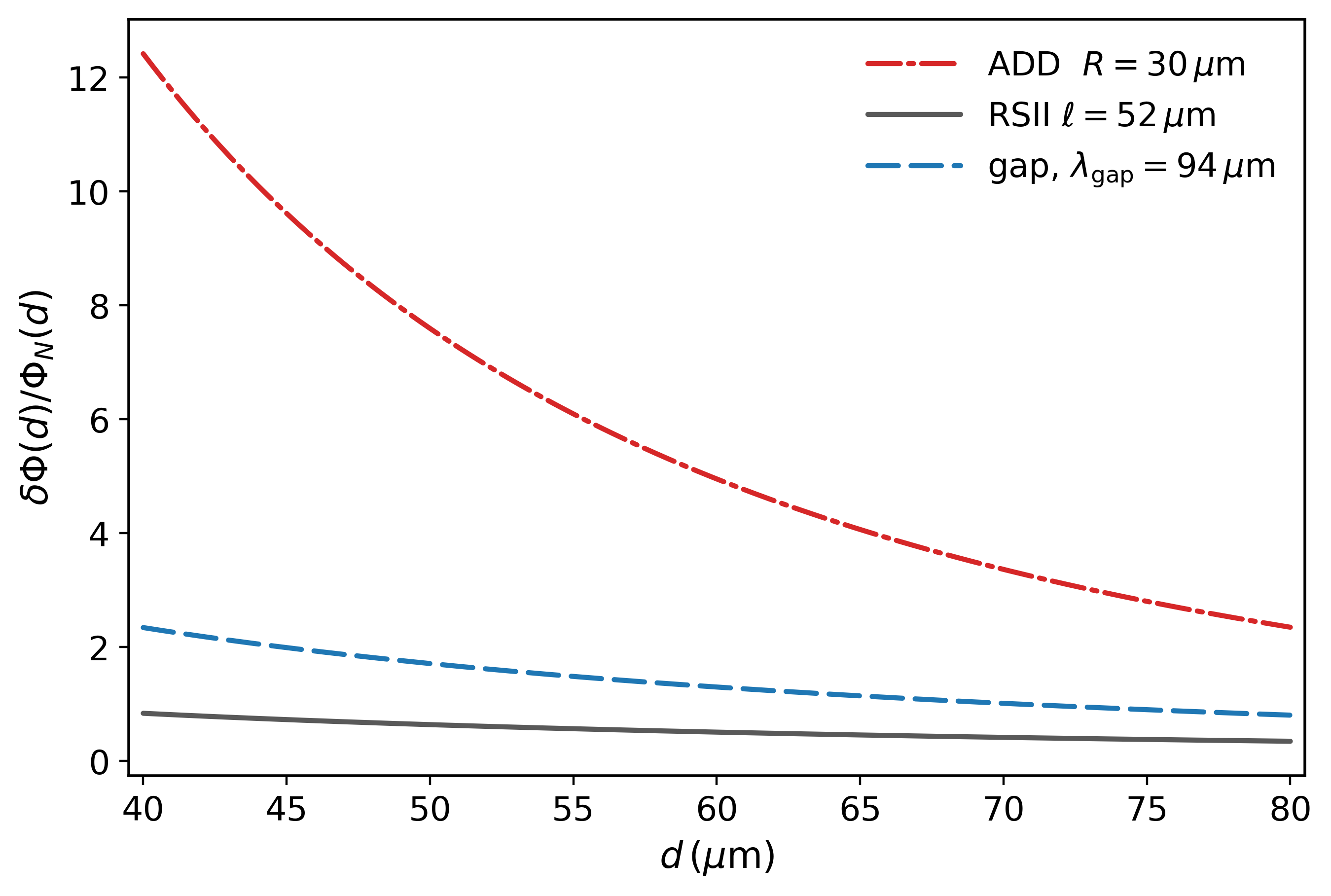}
\caption{Fractional entangling-phase shift $\delta\Phi(d)/\Phi_N(d)$ with the conservative benchmark~\eqref{eq:bench_cons}. Model parameters are ADD ($\delta=2$, $R=30\,\mu{\rm m}$), RSII ($\ell=52\,\mu{\rm m}$), and the gapped continuum ($\lambda_{\rm gap}=94\,\mu{\rm m}$). Over the plotted window, the benchmark ordering is ADD $>$ gapped $>$ RSII.}
\label{fig:phiratio_cons}
\end{figure}
	
Figure~\ref{fig:phiratio_cons} displays the fractional entangling-phase shift $\delta\Phi(d)/\Phi_N(d)$ for the three spectral classes. For all models, the fractional phase shift decreases as the separation increases, which reflects the suppression of short-range corrections at larger distances. At the smallest separation ($d\simeq 40\,\mu{\rm m}$), the ADD benchmark yields $\delta\Phi/\Phi_N\simeq 12$ while the gapped continuum and RSII benchmarks reach approximately 2.3 and 0.8, respectively. At the largest separation ($d\simeq 80\,\mu{\rm m}$), the same ordering persists, with ADD still of order unity while the gapped and RSII curves drop to roughly 0.8 and 0.35. The benchmark comparison thus reveals a stable hierarchy of ADD $>$ gapped $>$ RSII across the entire window, where the gapped continuum remains clearly distinguishable from RSII despite being far below the ADD signal.
		
The consistency between the hierarchy in Fig.~\ref{fig:amplification_factor} and the fractional phase shift in Fig.~\ref{fig:phiratio_cons} demonstrates that the amplification factor $\mathcal{A}(d)$ is the primary factor distinguishing these spectral classes. Although the fractional phase shift is determined by the product $\delta\Phi/\Phi_N \simeq \Xi(d)=\mathcal{A}(d)\Delta(d)$, the significantly stronger amplification factor $\mathcal{A}(d)$ in the gapped case ensures that its phase signature remains well-separated from the RSII baseline. Consequently, for the adopted benchmark choices, the gapped continuum and RSII models are not degenerate in a QGEM measurement.
 
\begin{figure}[tbp]
\centering
\includegraphics[width=\linewidth]{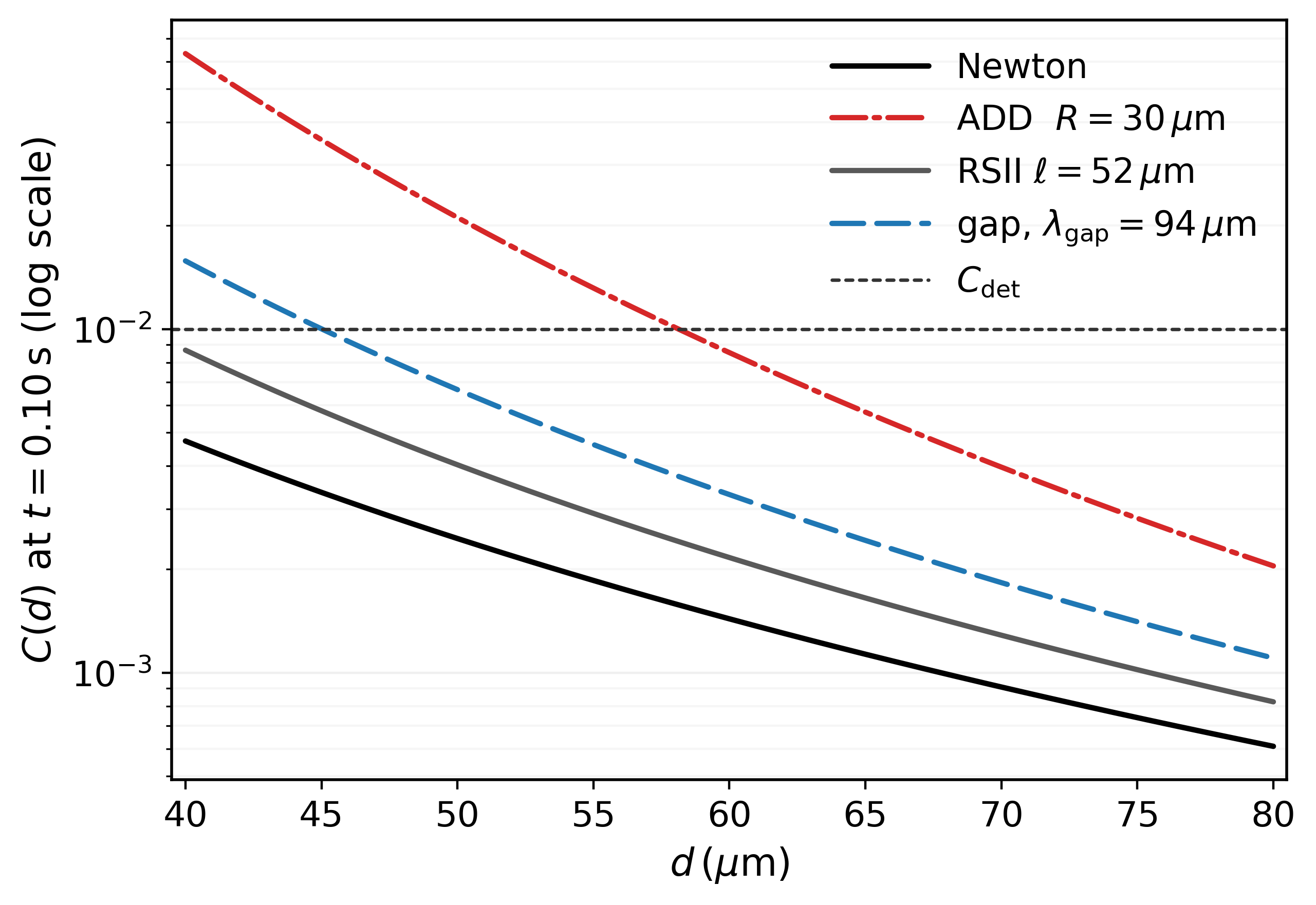}
\caption{Concurrence $C(d)$ for the conservative benchmark~\eqref{eq:bench_cons} on a logarithmic scale. The black dashed line indicates the detection threshold $C_{\rm det}=10^{-2}$, while the thin black line shows the Newtonian prediction for reference.}
\label{fig:conc_cons}
\end{figure}

The corresponding concurrence is presented in Fig.~\ref{fig:conc_cons}. We assume a minimal resolvable entangling phase of $\Phi_{\rm det}=0.02\,{\rm rad}$, which is consistent with the resolution thresholds of $\mathcal{O}(10^{-2})$ to $\mathcal{O}(10^{-1})\,\mathrm{rad}$ targeted in near-term QGEM proposals~\cite{vanDeKamp:2020PRA,Schut:2024Micrometer}. This choice leads to a detection threshold of $C_{\rm det}\simeq 10^{-2}$ via Eq.~\eqref{eq:Cdet_def_results}. In this conservative scenario, the standard Newtonian signal remains below the threshold throughout the entire window. Similarly, the RSII benchmark fails to cross the threshold at any point. The gapped benchmark represents an intermediate case that exceeds the threshold only at the smallest separations before dropping below it for the remainder of the scan. In contrast, the ADD benchmark remains detectable over a much broader interval. These results indicate that under conservative experimental conditions, the non-Newtonian benchmarks fall into three distinct categories, namely sustained threshold crossing for ADD, edge-of-window reach for the gapped continuum, and no detectable signal for RSII.

\subsection{Optimistic experimental parameters}
\label{subsec:opt_bench}
	
For a more optimistic near-term setting, we adopt the parameters
\begin{equation}
	m=2\times10^{-14}\,{\rm kg},\quad
			\Delta x=15\,\mu{\rm m},\quad
			t=0.30\,{\rm s},
\label{eq:bench_opt}
\end{equation}
which reflect the direction of proposed improvements in coherence time and background suppression in parallel geometries~\cite{vanDeKamp:2020PRA,Schut:2023PRR,Schut:2024Micrometer}. Since the fractional entangling-phase shift $\delta\Phi(d)/\Phi_N(d)$ is essentially independent of $(m,t)$ and depends only on the potential shape in the small-splitting regime, the optimistic fractional phase-shift curves are nearly identical to those in Fig.~\ref{fig:phiratio_cons} and are therefore not displayed.

\begin{figure}[tbp]
\centering
\includegraphics[width=\linewidth]{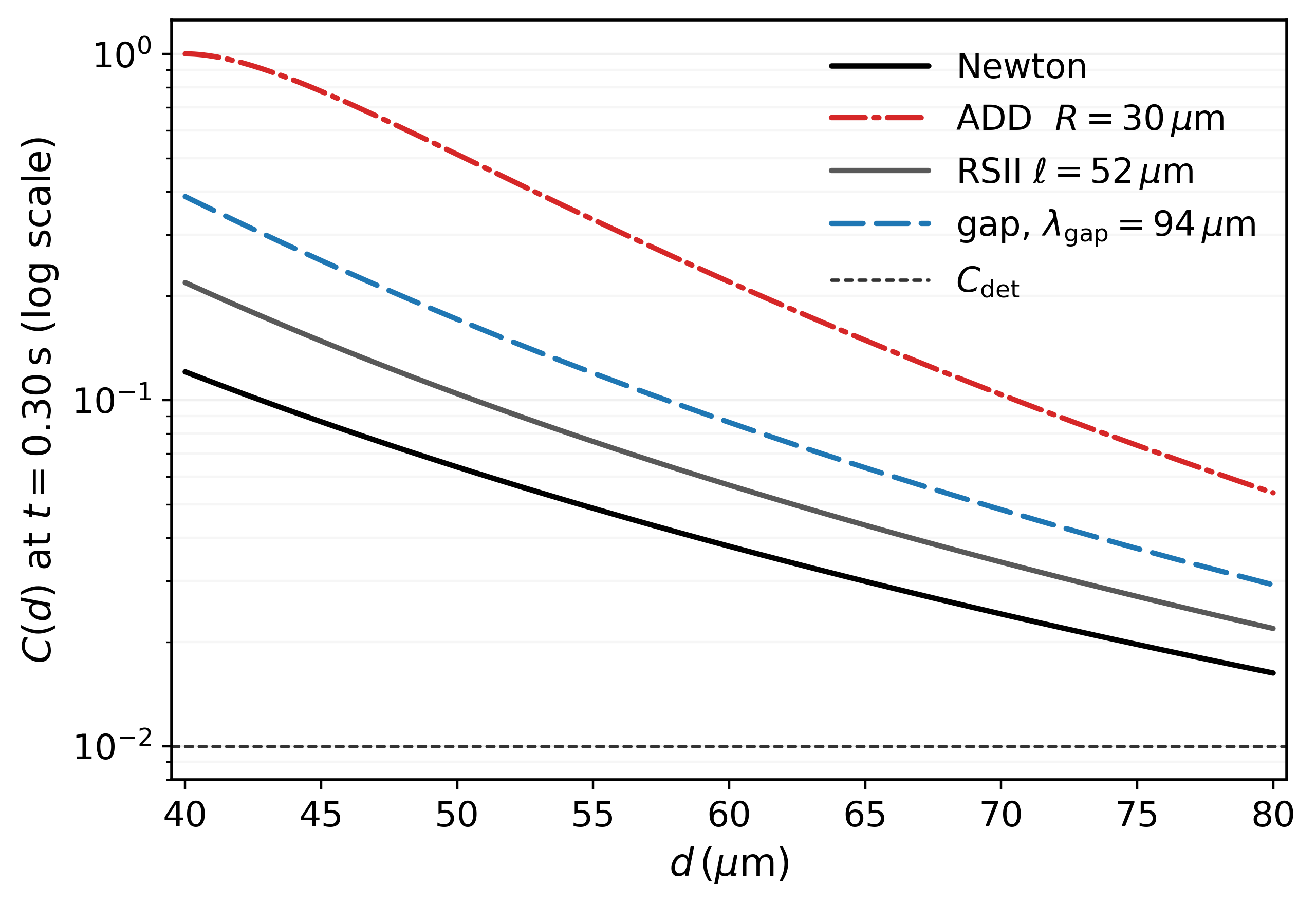}
\caption{Concurrence $C(d)$ for the optimistic benchmark~\eqref{eq:bench_opt} on a logarithmic scale. The black dashed line indicates the detection threshold $C_{\rm det}=10^{-2}$, while the thin black line shows the Newtonian prediction for reference.}
\label{fig:conc_opt}
\end{figure}

In contrast, the concurrence shown in Fig.~\ref{fig:conc_opt} increases substantially due to the larger mass and longer interaction time. For this optimistic case, the Newtonian prediction itself exceeds the detection threshold $C_{\rm det}$ across the entire range of separations. The benchmark hierarchy maintains the order ADD $>$ gapped $>$ RSII $>$ Newton, with the gapped curve remaining visibly distinct from the RSII baseline throughout the window. At the lower end of the scan, the ADD benchmark approaches saturation ($C\simeq 1$) while the remaining curves stay well below this limit.

In the small-phase regime where $C(d)\approx |\Phi(d)|/2$, varying the parameters $(m,\Delta x,t)$ primarily rescales the overall amplitude, which is roughly proportional to $t\,m^2\Delta x^2$ as implied by Eq.~\eqref{eq:Phi_smallDx}, without appreciably altering the $d$-dependence. Consequently, the two concurrence plots on a logarithmic scale retain similar shapes. The optimistic benchmark thus mainly shifts the absolute reach upward rather than modifying the relative separation pattern inherited from the fractional phase shift.

Collectively, Figs.~\ref{fig:phiratio_cons}--\ref{fig:conc_opt} contrast the three KK spectral classes by anchoring each model to its respective experimental limit or theoretical benchmark. This approach ensures a consistent basis for comparison, with the resulting divergence of the phase and concurrence curves directly reflecting the unique spectral response characteristic of each class.

\subsection{Distance-scan model discrimination}
\label{subsec:distance_scan}

Having discussed the absolute benchmark signals above, we now turn to a complementary question: how much model discrimination can be obtained from the separation dependence itself. In the small-splitting regime, the leading model dependence of the fractional phase shift is encoded in the phase-response profile $\Xi(d)$, as introduced in Eq.~\eqref{eq:Xi_def}. Consequently, a distance scan probes the functional form of the underlying KK correction rather than merely its overall amplitude.

This profile-based analysis is necessary for two reasons. First, the absolute phase scale depends on experimental variables such as masses, superposition size, interaction time, and shielding geometry, along with residual electromagnetic backgrounds \cite{vanDeKamp:2020PRA,Schut:2023PRR,Schut:2024Micrometer,Marshman:2022Tomography}. Such factors introduce substantial normalization uncertainties. Second, distinct models may exhibit similar signal magnitudes at certain distances, creating an accidental amplitude degeneracy. To extract the underlying functional dependence and remove these multiplicative uncertainties, we define the normalized phase-response profile
\begin{equation}
	\widehat{\Xi}(d;d_0)\equiv \frac{\Xi(d)}{\Xi(d_0)},
\label{eq:Xi_hat}
\end{equation}
where $d_0 = 50\,\mu\text{m}$ serves as a reference anchor within the experimental scan window. This normalization effectively maps the relative distance dependence to a common pivot, enabling a standardized comparison of the functional trends across different spectral classes. By construction, $\widehat{\Xi}(d;d_0)$ eliminates overall multiplicative normalization uncertainties, providing a robust metric to assess profile-level separation or near-degeneracy over the finite observation range.

\begin{figure}[tbp]
\centering
\includegraphics[width=1\linewidth]{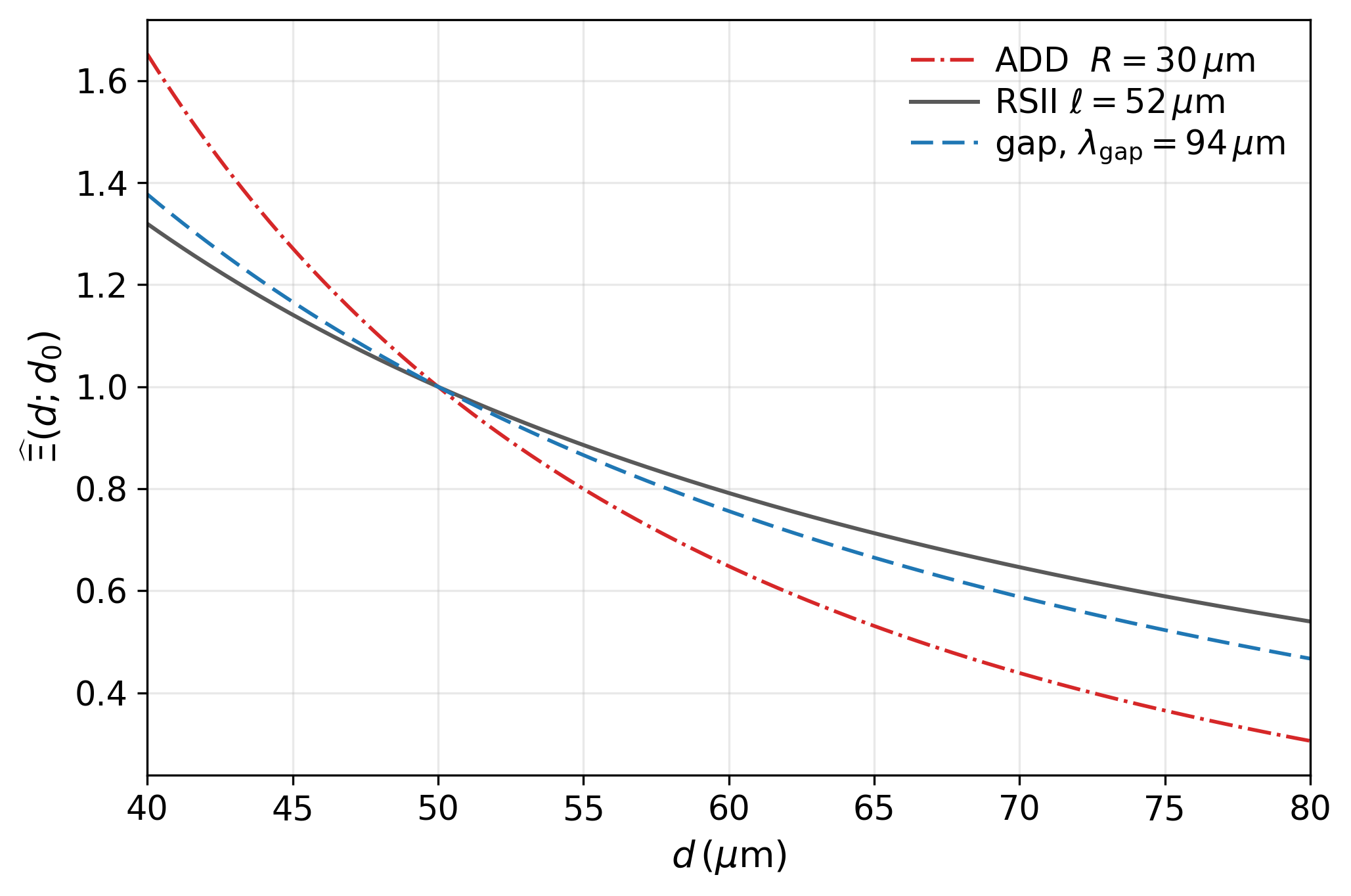}
\caption{Normalized phase-response profile $\widehat{\Xi}(d;d_0)$ with $d_0=50\,\mu\mathrm{m}$ in the scanning window $d=40$--$80\,\mu\mathrm{m}$ for the benchmark comparison used in the main text: ADD with $R=30\,\mu\mathrm{m}$, RSII with $\ell=52\,\mu\mathrm{m}$, and the gapped continuum with $\lambda_{\rm gap}=94\,\mu\mathrm{m}$.}
\label{fig:Xi_shape_near_degen}
\end{figure}

Figure~\ref{fig:Xi_shape_near_degen} shows the normalized benchmark curves for $R=30\,\mu{\rm m}$, $\ell=52\,\mu{\rm m}$, and $\lambda_{\rm gap}=94\,\mu{\rm m}$. Over the finite scan range, all curves intersect at $d=d_0$ by construction. Away from this point, the RSII curve exhibits the slowest decay, whereas the ADD curve decays most rapidly, with the gapped continuum situated in between. Consequently, distance scanning effectively distinguishes the RSII benchmark from the ADD and gapped scenarios. However, visual separation at the benchmark points does not by itself exclude the possibility of near-degeneracies elsewhere in parameter space.

\begin{figure}[tbp]
\centering
\includegraphics[width=\linewidth]{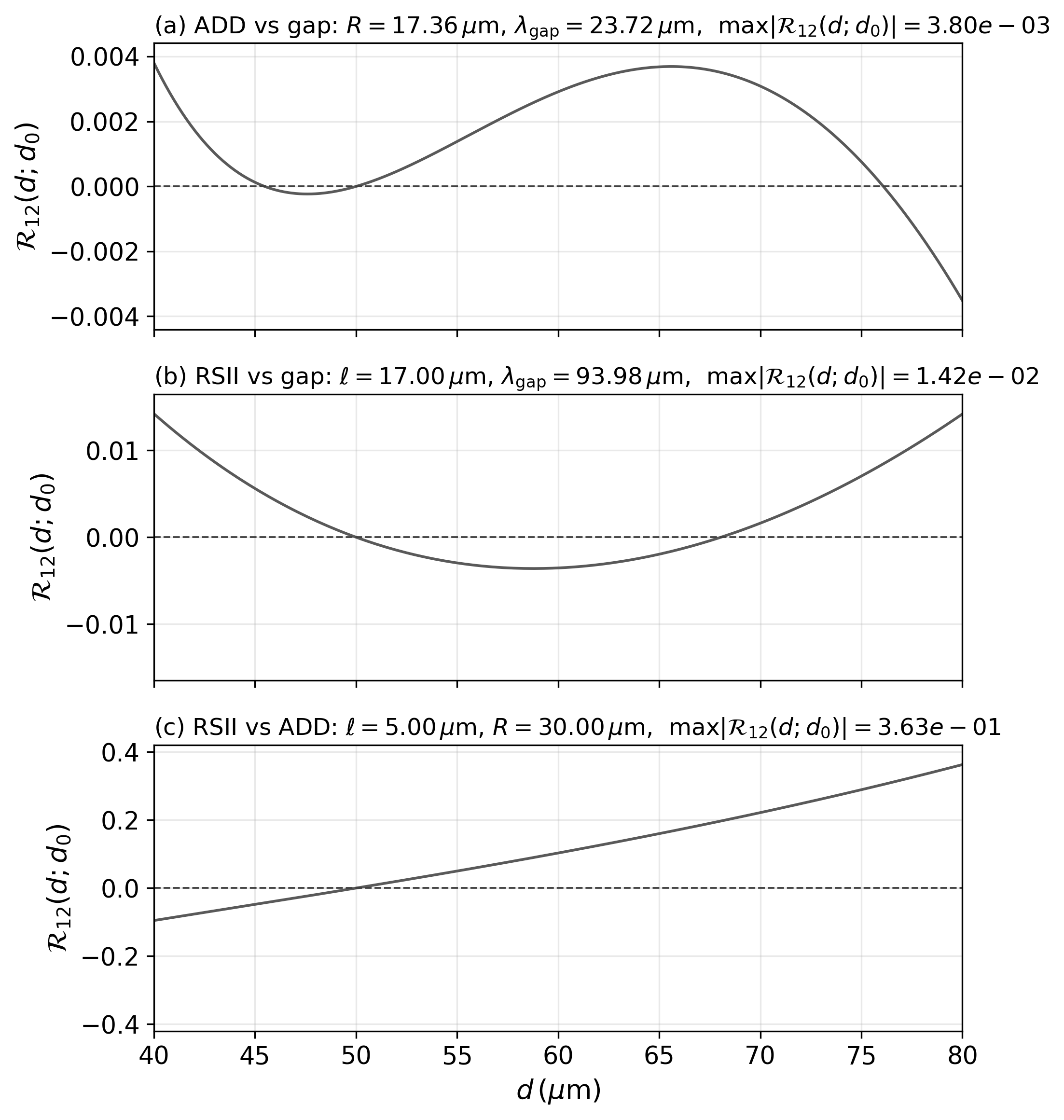}
\caption{Pairwise optimized normalized-profile residuals $\mathcal{R}_{12}(d;d_0)$ over the finite scan range $d=40$--$80\,\mu\mathrm{m}$. Panel (a) shows the optimized comparison between ADD and gapped continuum, with $(R,\lambda_{\rm gap})=(17.36,23.72)\,\mu\mathrm{m}$ and $\max|\mathcal{R}_{12}(d;d_0)| = 3.80\times10^{-3}$. Panel (b) shows the optimized comparison between RSII and gapped continuum, with $(\ell,\lambda_{\rm gap})=(17.00,93.98)\,\mu\mathrm{m}$ and $\max|\mathcal{R}_{12}(d;d_0)| = 1.42\times10^{-2}$. Panel (c) shows the optimized comparison between RSII and ADD, with $(\ell,R)=(5.00,30.00)\,\mu\mathrm{m}$ and $\max|\mathcal{R}_{12}(d;d_0)| = 3.63\times10^{-1}$.}
\label{fig:Xi_residual_near_degen}
\end{figure}

To quantify possible near-degeneracies in the normalized profile $\widehat{\Xi}(d;d_0)$, we compare pairs of normalized curves through the normalized-profile residual
\begin{equation}
	\mathcal{R}_{12}(d;d_0)\equiv
		\frac{\widehat{\Xi}_1(d;d_0)}{\widehat{\Xi}_2(d;d_0)}-1.
\label{eq:pairwise_residual}
\end{equation}
Here ``near-degeneracy'' refers specifically to the coincidence of the normalized profile curves $\widehat{\Xi}(d;d_0)$ over a finite scan range, rather than to equality of the absolute phase or concurrence amplitudes. For each model pair, we minimize the maximal residual over the finite scan range, i.e.,
\begin{equation}
\max_{d\in[40,80]\,\mu{\rm m}} |\mathcal{R}_{12}(d;d_0)|,
\end{equation}
with benchmark-motivated parameter ranges
\begin{equation}
	R \in [10,30]\,\mu{\rm m},\,
		\ell \in [5,52]\,\mu{\rm m},\,
		\lambda_{\rm gap} \in [10,94]\,\mu{\rm m}.
\end{equation}
The optimized residuals for each model pair are displayed in Fig.~\ref{fig:Xi_residual_near_degen}.

Our pairwise optimization analysis yields three key findings regarding model discriminability. First, as shown in Fig.~\ref{fig:Xi_residual_near_degen}(a), ADD and the gapped continuum can become nearly indistinguishable in normalized profile over the finite scan range. For the optimized pair $(R, \lambda_{\text{gap}}) = (17.36, 23.72)\,\mu\text{m}$, the maximum residual is suppressed to $\max|\mathcal{R}_{12}(d;d_0)| = 3.80 \times 10^{-3}$. Second, Fig.~\ref{fig:Xi_residual_near_degen}(b) reveals that RSII and the gapped continuum can reach moderate similarity but remain distinguishable at the percent level across the investigated ranges. In this case, the optimal pair $(\ell, \lambda_{\text{gap}}) = (17.00, 93.98)\,\mu\text{m}$ yields a residual of $1.42 \times 10^{-2}$. Finally, Fig.~\ref{fig:Xi_residual_near_degen}(c) reveals that RSII and ADD do not exhibit a comparable degree of degeneracy. Even for the most closely matched configuration $(\ell, R) = (5.00, 30.00)\,\mu\text{m}$, a substantial maximum residual of $3.63 \times 10^{-1}$ persists.

Therefore, the discrimination power of distance scanning is range-dependent. In the present benchmark comparison, $\widehat{\Xi}(d;d_0)$ cleanly separates the RSII benchmark curve from both ADD and the gapped benchmarks. By contrast, ADD and the gapped continuum can become nearly indistinguishable in normalized profile over a limited separation range. Consequently, breaking this degeneracy demands sub-percent profile precision, a broader separation range, or additional observables beyond the leading small-splitting response.

\section{Summary and Outlook}
\label{sec:Summary}

We have investigated the imprint of extra-dimensional corrections to Newtonian gravity at submillimeter separations in the QGEM framework, focusing on three representative KK spectral classes: ADD ($R=30\,\mu\mathrm{m}$), RSII ($\ell=52\,\mu{\rm m}$), and a PT-type gapped continuum ($\lambda_{\rm gap}=94\,\mu\mathrm{m}$). When evaluated at these benchmark limits, the QGEM phase and concurrence follow a consistent hierarchy: ADD $>$ gapped $>$ RSII. For conservative experimental parameters, the ADD signal exceeds the nominal detection threshold at smaller separations, whereas the gapped benchmark reaches it only at the lower edge of the scanning window. In contrast, the RSII signal remains below the resolution limit throughout the scan. For a more optimistic near-term setting, the extended experimental reach ensures that all three spectral benchmarks comfortably exceed the threshold, with the ADD signal reaching the strong-entanglement regime at minimal separations.
		
To complement these magnitude-based results, we introduce the normalized phase-response profile $\widehat{\Xi}(d;d_0)$ to evaluate the feasibility of distinguishing between models. Because $\widehat{\Xi}(d;d_0)$ is determined by the functional form of the gravitational correction rather than its absolute scale, it provides a diagnostic that is relatively insensitive to platform-specific normalization uncertainties. Our analysis shows that distance scanning effectively distinguishes the RSII profile from the other spectral classes. However, the ADD and gapped-continuum models exhibit a risk of phenomenological degeneracy, becoming nearly indistinguishable in normalized profile over a limited separation range. In such challenging cases, breaking the degeneracy would require sub-percent profile precision, an expanded scanning range, or the inclusion of higher-order observables beyond the leading-order small-splitting response.

Further refinement of this framework will involve the incorporation of experiment-specific geometries and comprehensive noise budgets, enabling rigorous likelihood analyses within the parameter spaces of specific braneworld configurations. It would also be of interest to extend this approach to more complex KK spectra, such as the resonance-dominated continuum, where $\Xi(d)$ and related profile diagnostics may exhibit distinctive crossover signatures. Such investigations will further clarify the utility of tabletop quantum sensors in probing the fundamental nature of gravity and the structure of extra dimensions.
	
\begin{acknowledgments}
We thank Dr.\ Wen-Tao Liu for helpful discussions on relativistic quantum information.
Y. Zhong acknowledges the support of the National Natural Science Foundation of China (Grant No.\ 12305061). 
K. Yang acknowledges the support of the National Natural Science Foundation of China (Grant No.~12475062) and the Natural Science Foundation of Chongqing (Grant No.~CSTB2024NSCQ-MSX0358).
\end{acknowledgments}


\begin{thebibliography}{10}

\bibitem{Randall:1999vf}
L.~Randall and R.~Sundrum, 
\emph{An alternative to compactification}, 
\href{https://doi.org/10.1103/PhysRevLett.83.4690}
{{Phys. Rev. Lett.} {\bfseries 83} (1999) 4690} 
[\href{https://arxiv.org/abs/hep-th/9906064}{{\ttfamily arXiv:hep-th/9906064}}].

\bibitem{Garriga:1999yh}
J.~Garriga and T.~Tanaka, 
\emph{Gravity in the randall--sundrum brane world}, 
\href{https://doi.org/10.1103/PhysRevLett.84.2778}
{{Phys. Rev. Lett.} {\bfseries 84} (2000) 2778} 
[\href{https://arxiv.org/abs/hep-th/9911055}{{\ttfamily arXiv:hep-th/9911055}}].

\bibitem{Callin:2004py}
P.~Callin and F.~Ravndal, 
\emph{Higher order corrections to the newtonian potential in the randall--sundrum model}, 
\href{https://doi.org/10.1103/PhysRevD.70.104009}
{{Phys. Rev. D} {\bfseries 70} (2004) 104009} 
[\href{https://arxiv.org/abs/hep-ph/0403302}{{\ttfamily arXiv:hep-ph/0403302}}].

\bibitem{Antoniadis:1998ig}
I.~Antoniadis, N.~Arkani-Hamed, S.~Dimopoulos and G.~Dvali, 
\emph{New dimensions at a millimeter to a fermi and superstrings at a tev}, 
\href{https://doi.org/10.1016/S0370-2693(98)00860-0}
{{Phys. Lett. B} {\bfseries 436} (1998) 257} 
[\href{https://arxiv.org/abs/hep-ph/9804398}{{\ttfamily arXiv:hep-ph/9804398}}].

\bibitem{ArkaniHamed:1998rs}
N.~Arkani-Hamed, S.~Dimopoulos and G.~Dvali, 
\emph{The hierarchy problem and new dimensions at a millimeter}, 
\href{https://doi.org/10.1016/S0370-2693(98)00466-3}
{{Phys. Lett. B} {\bfseries 429} (1998) 263} 
[\href{https://arxiv.org/abs/hep-ph/9803315}{{\ttfamily arXiv:hep-ph/9803315}}].

\bibitem{Gremm:2000dj}
M.~Gremm, 
\emph{Four-dimensional gravity on a thick domain wall}, 
\href{https://doi.org/10.1016/S0370-2693(00)00303-8}
{{Phys. Lett. B} {\bfseries 478} (2000) 434} 
[\href{https://arxiv.org/abs/hep-th/9912060}{{\ttfamily arXiv:hep-th/9912060}}].

\bibitem{BarbosaCendejas:2007MassGap}
N.~{Barbosa-Cendejas}, A.~{Herrera-Aguilar}, U.~Nucamendi, I.~Quiros and K.~Kanakoglou,
\emph{Mass hierarchy, mass gap and corrections to {Newton}'s law on thick branes with {Poincar{\'e}} symmetry}, 
\href{https://doi.org/10.1007/s10714-013-1631-9}
{{Gen. Rel. Grav.} {\bfseries 46} (2014) 1631} 
[\href{https://arxiv.org/abs/0712.3098}{{\ttfamily arXiv:0712.3098}}].

\bibitem{BarbosaCendejas:2008WeylMassGap}
N.~Barbosa-Cendejas, A.~Herrera-Aguilar, M.A.~Reyes~Santos and C.~Schubert, 
\emph{Mass gap for gravity localized on weyl thick branes}, 
\href{https://doi.org/10.1103/PhysRevD.77.126013}
{{Phys. Rev. D} {\bfseries 77} (2008) 126013} 
[\href{https://arxiv.org/abs/0709.3552}{{\ttfamily arXiv:0709.3552}}].

\bibitem{Sui:2021mimetic}
T.-T.~Sui, Y.-P.~Zhang, B.-M.~Gu and Y.-X.~Liu, 
\emph{Fundamental energy scale of the thick brane in mimetic gravity}, 
\href{https://doi.org/10.1140/epjc/s10052-021-09756-8}
{{Eur. Phys. J. C} {\bfseries 81} (2021) 980} 
[\href{https://arxiv.org/abs/2005.08438}{{\ttfamily arXiv:2005.08438}}].

\bibitem{Adelberger:2009zz}
E.G.~Adelberger, J.H.~Gundlach, B.R.~Heckel, S.~Hoedl and S.~Schlamminger, 
\emph{Torsion balance experiments: A low-energy frontier of particle physics}, 
\href{https://doi.org/10.1016/j.ppnp.2008.08.002}
{{Prog. Part. Nucl. Phys.} {\bfseries 62} (2009) 102}.

\bibitem{Kapner:2006si}
D.J.~Kapner, T.S.~Cook, E.G.~Adelberger, J.H.~Gundlach, B.R.~Heckel, C.D.~Hoyle et~al., 
\emph{Tests of the gravitational inverse-square law below the dark-energy length scale}, 
\href{https://doi.org/10.1103/PhysRevLett.98.021101}
{{Phys. Rev. Lett.} {\bfseries 98} (2007) 021101} 
[\href{https://arxiv.org/abs/hep-ph/0611184}{{\ttfamily arXiv:hep-ph/0611184}}].

\bibitem{Tan:2016}
W.-H.~Tan, A.-B.~Du, W.-C.~Dong, S.-Q.~Yang, C.-G.~Shao, S.-G.~Guan et~al., 
\emph{New test of the gravitational inverse-square law at the submillimeter range}, 
\href{https://doi.org/10.1103/PhysRevLett.116.131101}
{{Phys. Rev. Lett.} {\bfseries 116} (2016) 131101}.

\bibitem{Lee:2020}
J.G.~Lee, E.G.~Adelberger, T.S.~Cook, S.M.~Fleischer and B.R.~Heckel, 
\emph{New test of the gravitational $1/r^2$ law at separations down to $52\,\mu$m}, 
\href{https://doi.org/10.1103/PhysRevLett.124.101101}
{{Phys. Rev. Lett.} {\bfseries 124} (2020) 101101} 
[\href{https://arxiv.org/abs/2002.11761}{{\ttfamily arXiv:2002.11761}}].

\bibitem{Tan:2020}
W.-H.~Tan, A.-B.~Du, W.-C.~Dong, S.-Q.~Yang, C.-G.~Shao, S.-G.~Guan et~al., 
\emph{Improvement for testing the gravitational inverse-square law at the submillimeter range}, 
\href{https://doi.org/10.1103/PhysRevLett.124.051301}
{{Phys. Rev. Lett.} {\bfseries 124} (2020) 051301}.

\bibitem{Klimchitskaya:2009RMP}
G.L.~Klimchitskaya, U.~Mohideen and V.M.~Mostepanenko, 
\emph{The casimir force between real materials: Experiment and theory}, 
\href{https://doi.org/10.1103/RevModPhys.81.1827}
{{Rev. Mod. Phys.} {\bfseries 81} (2009) 1827} 
[\href{https://arxiv.org/abs/0902.4022}{{\ttfamily arXiv:0902.4022}}].

\bibitem{Behunin:2014Patch}
R.O.~Behunin, D.A.R.~Dalvit, R.S.~Decca and C.C.~Speake, 
\emph{Limits on the accuracy of force sensing at short separations due to patch potentials}, 
\href{https://doi.org/10.1103/PhysRevD.89.051301}
{{Phys. Rev. D} {\bfseries 89} (2014) 051301} 
[\href{https://arxiv.org/abs/1304.4074}{{\ttfamily arXiv:1304.4074}}].

\bibitem{Bose:2017SpinWitness}
S.~Bose, A.~Mazumdar, G.W.~Morley, H.~Ulbricht, M.~Toro{\v s}, M.~Paternostro et~al., 
\emph{Spin entanglement witness for quantum gravity}, 
\href{https://doi.org/10.1103/PhysRevLett.119.240401}
{{Phys. Rev. Lett.} {\bfseries 119} (2017) 240401} 
[\href{https://arxiv.org/abs/1707.06050}{{\ttfamily arXiv:1707.06050}}].

\bibitem{MarlettoVedral:2017BMV}
C.~Marletto and V.~Vedral, 
\emph{Gravitationally induced entanglement between two massive particles is sufficient evidence of quantum effects in gravity}, 
\href{https://doi.org/10.1103/PhysRevLett.119.240402}
{{Phys. Rev. Lett.} {\bfseries 119} (2017) 240402} 
[\href{https://arxiv.org/abs/1707.06036}{{\ttfamily arXiv:1707.06036}}].

\bibitem{MarlettoVedral:2025RMP}
C.~Marletto and V.~Vedral, 
\emph{Quantum-information methods for quantum gravity laboratory-based tests}, 
\href{https://doi.org/10.1103/RevModPhys.97.015006}
{{Rev. Mod. Phys.} {\bfseries 97} (2025) 015006} 
[\href{https://arxiv.org/abs/2410.07262}{{\ttfamily arXiv:2410.07262}}].

\bibitem{Bose:2025Interfaces}
S.~Bose, I.~Fuentes, A.A.~Geraci, S.M.~Khan, S.~Qvarfort, M.~Rademacher et~al., 
\emph{Massive quantum systems as interfaces of quantum mechanics and gravity}, 
\href{https://doi.org/10.1103/RevModPhys.97.015003}
{{Rev. Mod. Phys.} {\bfseries 97} (2025) 015003} 
[\href{https://arxiv.org/abs/2311.09218}{{\ttfamily arXiv:2311.09218}}].

\bibitem{KafriTaylorMilburn:2014ClassicalChannel}
D.~Kafri, J.M.~Taylor and G.J.~Milburn, 
\emph{A classical channel model for gravitational decoherence}, 
\href{https://doi.org/10.1088/1367-2630/16/6/065020}
{{New J. Phys.} {\bfseries 16} (2014) 065020} 
[\href{https://arxiv.org/abs/1401.0946}{{\ttfamily arXiv:1401.0946}}].

\bibitem{TilloyDiosi:2017LeastDecoherence}
A.~Tilloy and L.~Di{\'o}si, 
\emph{Principle of least decoherence for newtonian semi-classical gravity}, 
\href{https://doi.org/10.1103/PhysRevD.96.104045}
{{Phys. Rev. D} {\bfseries 96} (2017) 104045} 
[\href{https://arxiv.org/abs/1706.01856}{{\ttfamily arXiv:1706.01856}}].

\bibitem{AnastopoulosHu:2022Entropy}
C.~Anastopoulos and B.-L.~Hu, 
\emph{Gravity, quantum fields and quantum information: Problems with classical channel and stochastic theories}, 
\href{https://doi.org/10.3390/e24040490}
{{Entropy} {\bfseries 24} (2022) 490} 
[\href{https://arxiv.org/abs/2202.02789}{{\ttfamily arXiv:2202.02789}}].

\bibitem{AzizHowl:2025Nature}
J.~Aziz and R.~Howl, 
\emph{Classical theories of gravity produce entanglement}, 
\href{https://doi.org/10.1038/s41586-025-09595-7}
{{Nature} {\bfseries 646} (2025) 813} 
[\href{https://arxiv.org/abs/2510.19714}{{\ttfamily arXiv:2510.19714}}].

\bibitem{Chevalier:2020UnknownInteractions}
H.~Chevalier, A.J.~Paige and M.S.~Kim, 
\emph{Witnessing the non-classical nature of gravity in the presence of unknown interactions}, 
\href{https://doi.org/10.1103/PhysRevA.102.022428}
{{Phys. Rev. A} {\bfseries 102} (2020) 022428} 
[\href{https://arxiv.org/abs/2005.13922}{{\ttfamily arXiv:2005.13922}}].

\bibitem{MarlettoVedral:2025LocalMeans}
C.~Marletto and V.~Vedral, 
\emph{Classical gravity cannot mediate entanglement by local means}, 2025.

\bibitem{MarlettoOppenheimVedralWilson:2025ClassicalCannot}
C.~Marletto, J.~Oppenheim, V.~Vedral and E.~Wilson, 
\emph{Classical gravity cannot mediate entanglement}, 2025.

\bibitem{Diosi:2025NoClassicalGravityEntangles}
L.~Di{\'o}si, \emph{No, classical gravity does not entangle quantized matter fields},  2025.

\bibitem{Marshman:2022Tomography}
P.F.~Barker, S.~Bose, R.J.~Marshman and A.~Mazumdar, 
\emph{Entanglement based tomography to probe new macroscopic forces}, 
\href{https://doi.org/10.1103/PhysRevD.106.L041901}
{{Phys. Rev. D} {\bfseries 106} (2022) L041901} 
[\href{https://arxiv.org/abs/2203.00038}{{\ttfamily arXiv:2203.00038}}].

\bibitem{CarmonaRufo:2025ALPWitness}
P.G.~Carmona~Rufo, A.~Kumar, C.~Sab{\'i}n and A.~Mazumdar, 
\emph{Entanglement witnesses mediated via axionlike particles}, 
{{Phys. Rev. D} {\bfseries 111} (2025) 115005} 
[\href{https://arxiv.org/abs/2503.19072}{{\ttfamily arXiv:2503.19072}}].

\bibitem{Elahi:2023ProbingMassive}
S.G.~Elahi and A.~Mazumdar, 
\emph{Probing massless and massive gravitons via entanglement in a warped extra dimension}, 
\href{https://doi.org/10.1103/PhysRevD.108.035018}
{{Phys. Rev. D} {\bfseries 108} (2023) 035018} 
[\href{https://arxiv.org/abs/2303.07371}{{\ttfamily arXiv:2303.07371}}].

\bibitem{Feng:2024QGEMExtraDim}
S.~Feng, B.-M.~Gu and F.-W.~Shu, 
\emph{Quantum gravity induced entanglement of masses with extra dimensions}, 
\href{https://doi.org/10.1140/epjc/s10052-024-12413-5}
{{Eur. Phys. J. C} {\bfseries 84} (2024) 59} 
[\href{https://arxiv.org/abs/2307.11391}{{\ttfamily arXiv:2307.11391}}].

\bibitem{vanDeKamp:2020PRA}
T.W.~van~de Kamp, R.J.~Marshman, S.~Bose and A.~Mazumdar, 
\emph{Quantum gravity witness via entanglement of masses: Casimir screening}, \href{https://doi.org/10.1103/PhysRevA.102.062807}
{{Phys. Rev. A} {\bfseries 102} (2020) 062807} 
[\href{https://arxiv.org/abs/2006.06931}{{\ttfamily arXiv:2006.06931}}].

\bibitem{Schut:2023PRR}
M.~Schut, A.~Grinin, A.~Dana, S.~Bose, A.~Geraci and A.~Mazumdar, 
\emph{Relaxation of experimental parameters in a quantum-gravity-induced entanglement of masses protocol using electromagnetic screening}, 
\href{https://doi.org/10.1103/PhysRevResearch.5.043170}
{{Phys. Rev. Res.} {\bfseries 5} (2023) 043170} 
[\href{https://arxiv.org/abs/2307.07536}{{\ttfamily arXiv:2307.07536}}].

\bibitem{Schut:2024Micrometer}
M.~Schut, A.~Geraci, S.~Bose and A.~Mazumdar, 
\emph{Micrometer-size spatial superpositions for the {QGEM} protocol via screening and trapping}, 
\href{https://doi.org/10.1103/PhysRevResearch.6.013199}
{{Phys. Rev. Res.} {\bfseries 6} (2024) 013199} 
[\href{https://arxiv.org/abs/2307.15743}{{\ttfamily arXiv:2307.15743}}].

\bibitem{Csaki:2000fc}
C.~Csaki, J.~Erlich, T.J.~Hollowood and Y.~Shirman, 
\emph{Universal aspects of gravity localized on thick branes}, 
\href{https://doi.org/10.1016/S0550-3213(00)00271-6}
{{Nucl. Phys. B} {\bfseries 581} (2000) 309} 
[\href{https://arxiv.org/abs/hep-th/0001033}{{\ttfamily arXiv:hep-th/0001033}}].

\bibitem{Navas:2024RPP}
S.~Navas, others and {Particle Data Group}, 
\emph{Review of particle physics}, 
\href{https://doi.org/10.1103/PhysRevD.110.030001}
{{Phys. Rev. D} {\bfseries 110} (2024) 030001}.

\bibitem{Nam:1999MassGapKK}
S.~Nam, 
\emph{Mass gap in kaluza-klein spectrum in a network of brane worlds}, \href{https://doi.org/10.1088/1126-6708/2000/04/002}
{{JHEP} {\bfseries 04} (2000) 002} 
[\href{https://arxiv.org/abs/hep-th/9911237}{{\ttfamily arXiv:hep-th/9911237}}].

\end{thebibliography}
\end{document}